\documentclass[12pt,english]{article}
\usepackage[T1]{fontenc}
\usepackage[latin9]{inputenc}
\usepackage{verbatim}
\usepackage{refstyle}
\usepackage{amsmath}
\usepackage{graphicx}
\usepackage[numbers,super,sort&compress]{natbib}

\makeatletter

\AtBeginDocument{\providecommand\secref[1]{\ref{sec:#1}}}
\AtBeginDocument{\providecommand\eqref[1]{\ref{eq:#1}}}
\AtBeginDocument{\providecommand\subsecref[1]{\ref{subsec:#1}}}
\AtBeginDocument{\providecommand\figref[1]{\ref{fig:#1}}}
\RS@ifundefined{subsecref}
  {\newref{subsec}{name = \RSsectxt}}
  {}
\RS@ifundefined{thmref}
  {\def\RSthmtxt{theorem~}\newref{thm}{name = \RSthmtxt}}
  {}
\RS@ifundefined{lemref}
  {\def\RSlemtxt{lemma~}\newref{lem}{name = \RSlemtxt}}
  {}

\newcommand{\lyxaddress}[1]{
	\par {\raggedright #1
	\vspace{1.4em}
	\noindent\par}
}
\newcommand{\lyxrightaddress}[1]{
	\par {\raggedleft \begin{tabular}{l}\ignorespaces
	#1
	\end{tabular}
	\vspace{1.4em}
	\par}
}

\usepackage{braket}
\usepackage[version=3]{mhchem}
\usepackage{url}
\usepackage[shortcuts]{extdash}

\newref{sec}{name = section~, names = sections~, Name = Section~, Names = Sections~}
\newref{subsec}{name = section~, names = sections~}

\@ifundefined{showcaptionsetup}{}{%
 \PassOptionsToPackage{caption=false}{subfig}}
\usepackage{subfig}
\makeatother

\usepackage{babel}
\begin{document}
\title{A review on non-relativistic, fully numerical electronic structure
calculations on atoms and diatomic molecules}
\author{Susi Lehtola}
\maketitle

\lyxaddress{Department of Chemistry, University of Helsinki, P.O. Box 55 (A.
I. Virtasen aukio 1), FI-00014 University of Helsinki, Finland}

\lyxrightaddress{susi.lehtola@alumni.helsinki.fi}
\begin{abstract}
The need for accurate calculations on atoms and diatomic molecules
is motivated by the opportunities and challenges of such studies.
The most commonly-used approach for all-electron electronic structure
calculations in general -- the linear combination of atomic orbitals
(LCAO) method -- is discussed in combination with Gaussian, Slater
a.k.a.\emph{ }exponential, and numerical radial functions. Even though
LCAO calculations have major benefits, their shortcomings motivate
the need for fully numerical approaches based on, e.g. finite differences,
finite elements, or the discrete variable representation, which are
also briefly introduced.

Applications of fully numerical approaches for general molecules are
briefly reviewed, and their challenges are discussed. It is pointed
out that the high level of symmetry present in atoms and diatomic
molecules can be exploited to fashion more efficient fully numerical
approaches for these special cases, after which it is possible to
routinely perform all-electron Hartree--Fock and density functional
calculations directly at the basis set limit on such systems. Applications
of fully numerical approaches to calculations on atoms as well as
diatomic molecules are reviewed. Finally, a summary and outlook is
given.
\end{abstract}
\global\long\def\ERI#1#2{(#1|#2)}%
\global\long\def\bra#1{\Bra{#1}}%
\global\long\def\ket#1{\Ket{#1}}%
\global\long\def\braket#1{\Braket{#1}}%

\newcommand*\citeref[1]{ref. \citenum{#1}}
\newcommand*\citerefs[1]{refs. \citenum{#1}}

\section{Introduction\label{sec:Introduction}}

Thanks to decades of development in approximate density functionals
and numerical algorithms, density functional theory\citep{Hohenberg1964,Kohn1965}
(DFT) is now one of the cornerstones of computational chemistry and
solid state physics.\citep{Becke2014,Jones2015,Mardirossian2017a}
Since atoms are the basic building block of molecules, a number of
popular density functionals\citep{Becke1986a,Becke1988a,Becke1997,Sun2015,Sun2015a}
have been parametrized using \emph{ab initio} data on noble gas atoms;
these functionals have been used in turn as the starting point for
the development of other functionals. A comparison of the results
of density-functional calculations to accurate \emph{ab initio }data
on atoms has, however, recently sparked controversy on the accuracy
of a number of recently published, commonly-used density functionals.\citep{Medvedev2017,Brorsen2017,Kepp2017,Medvedev2017a,Hammes-Schiffer2017,Korth2017,Gould2017,Mezei2017,Wang2017a,Kepp2018,Su2018}
This underlines that there is still room for improvement in DFT even
for the relatively simple case of atomic studies.

The development and characterization of new density functionals requires
the ability to perform accurate atomic calculations. Because atoms
are the simplest chemical systems, atomic calculations may seem simple;
however, they are in fact sometimes surprisingly challenging. For
instance, a long-standing discrepancy between the theoretical and
experimental electron affinity and ionization potential of gold has
been resolved only very recently with high-level \emph{ab initio}
calculations.\citep{Pasteka2017} Atomic calculations can also be
used to formulate efficient starting guesses for molecular electronic
structure calculations via either the atomic density\citep{Almlof1982,VanLenthe2006}
or the atomic potential as discussed in \citeref{Lehtola2019}.

Diatomic molecules, in turn, again despite their apparent simplicity
offer a profound richness of chemistry; ranging from the simple covalent
bond in \ce{H2} to the sextuple bonds in \ce{Cr2}, \ce{Mo2}, and
\ce{W2};\citep{Roos2007} and from the strong ionic bond in \ce{Na+Cl-}
to the weak dispersion interactions in the noble gases such as \ce{Ne\bond{...}Ar}.
Even the deceivingly simple \ce{Be2} molecule is a challenge to many
theoretical methods, due to its richness of static and dynamic correlation
effects.\citep{Karton2018} Counterintuitive non-nuclear electron
density maxima can also be found in diatomic molecules, both in the
homonuclear and heteronuclear case,\citep{Terrabuio2013} again highlighting
the surprising richness in the electronic structure of diatomic molecules.
Going further down in the periodic table, diatomic molecules can be
used to study relativistic effects to chemical bonding\citep{Pyykko2012a,Pyykko2012}
as well as the accuracy of various relativistic approaches.\citep{Desclaux1976,Pyykko1976,Pyykko1979,Sepp1986,Bastug1993,Pyykko1979a,Snijders1980,Park1994,VanLenthe1994,Visscher1996,Bastug1997,Liu1997a,DeJong1998,Nakajima1999,Nakajima1999a,Faegri2001,Anton2002,Zhang2004,Roos2006,Hofener2012,Wang2016c,Knecht2018}
Although much of the present discussion applies straightforwardly
to relativistic methodologies as well, they will not be discussed
further in this manuscript. Instead, in the present work, I will review
basic approaches for modeling the non-relativistic electronic structure
of atoms and diatomic molecules to high accuracy \emph{i.e. }close
to the complete basis set (CBS) limit.

The organisation of the manuscript is the following. In \secref{LCAO-approaches},
I will give an overview of commonly-used linear combination of atomic
orbitals (LCAO) approaches. I will discuss both the advantages as
well as the deficiencies of LCAO methods, the latter of which motivate
the need for fully numerical alternatives, which are then summarized
in \secref{Real-space-methods}. \Secref{Applications} discusses
applications of fully numerical approaches, starting out with a summary
review on approaches for general molecules in \subsecref{Overview-on-general},
and continuing with a detailed review on the special approaches for
atoms and diatomic molecules in \subsecref{Atoms, Diatomic-molecules},
respectively. A summary and outlook is given in \secref{Summary-and-outlook}.
Atomic units are used throughout the work.

\section{LCAO approaches\label{sec:LCAO-approaches}}

In order to perform an electronic structure calculation, one must
first specify the degrees of freedom that are allowed for the one-particle
states a.k.a. orbitals of the many-electron wave function; that is,
a basis set must be introduced. An overwhelming majority of the all-electron
quantum chemical calculations that have been, and still are being
reported in the literature employ basis sets constructed of atomic
orbitals (AOs) through the LCAO approach.\citep{Heitler1927,Lennard-Jones1929}
As the name suggests, in LCAO the electronic single-particle states
\emph{i.e.} molecular orbitals $\psi_{i}^{\sigma}$ of spin $\sigma$
are expanded in terms of AO basis functions as
\begin{equation}
\psi_{i}^{\sigma}(\boldsymbol{r})=\sum_{\alpha}C_{\alpha i}^{\sigma}\chi_{n_{\alpha}l_{\alpha}m_{\alpha}}(\boldsymbol{r}_{\alpha}),\label{eq:LCAO}
\end{equation}
where the index $\alpha$ runs over all shells of all atoms in the
system, $\boldsymbol{r}_{\alpha}$ denotes the distance $\boldsymbol{r}_{\alpha}=\boldsymbol{r}-\boldsymbol{R}_{\alpha}$,
and $\boldsymbol{R}_{\alpha}$ is the center of the $\alpha^{\text{th}}$
basis function that typically coincides with a nucleus. The matrix
$\boldsymbol{C}^{\sigma}$ contains the molecular orbital coefficients,
its rows and columns corresponding to the basis function and molecular
orbital indices, respectively. The quantities $n$, $l$, and $m$
in \eqref{LCAO} are the principal, azimuthal, and magnetic quantum
numbers, respectively, of the AO basis functions
\begin{equation}
\chi_{nlm}(\boldsymbol{r})=R_{nl}(r)Y_{l}^{m}(\hat{\boldsymbol{r}}),\label{eq:atorb}
\end{equation}
where $R_{nl}(r)$ are radial functions, choices for which are discussed
below in \subsecref{Variants-of-LCAO}, and $Y_{l}^{m}(\hat{\boldsymbol{r}})$
are spherical harmonics that are usually chosen in the real form.

The variational solution of the Hartree--Fock (HF) or Kohn--Sham\citep{Kohn1965}
equations within the LCAO basis set leads to the Roothaan--Hall or
Pople--Nesbet equations\citep{Roothaan1951,Pople1954,Pople1992}
for the molecular orbitals
\begin{equation}
\boldsymbol{F}^{\sigma}\boldsymbol{C}^{\sigma}=\boldsymbol{S}\boldsymbol{C}^{\sigma}\boldsymbol{E}^{\sigma},\label{eq:Roothaan}
\end{equation}
where $\boldsymbol{F}^{\sigma}=\boldsymbol{F}^{\sigma}(\boldsymbol{C}^{\alpha},\boldsymbol{C}^{\beta})$
is the (Kohn--Sham--)Fock matrix, $\boldsymbol{S}$ is the overlap
matrix, and $\boldsymbol{E}^{\sigma}$ is a diagonal matrix containing
the orbital energies. Due to the interdependence of the Fock matrix
and the molecular orbitals, \eqref{Roothaan} has to be solved self-consistently,
leading to the self-consistent field (SCF) approach. The iterative
solution of \eqref{Roothaan} starts from an initial guess for the
orbitals, choices for which I have reviewed recently in \citeref{Lehtola2019}.

\subsection{Benefits of LCAO approaches}

LCAO calculations allow all electrons to be explicitly included in
the calculation -- even if it includes heavy atoms -- without significant
additional computational cost, as the necessary number of basis functions
scales only in proportion to the number of electrons; the prefactor
furthermore being very small. Moreover, since the AO basis functions
tend to look a lot like real AOs, LCAO basis set truncation errors
are often systematic, \emph{i.e.} the error is similar at different
geometries or spin states, for example. In such a case, even if the
errors themselves are large, the relative error between different
geometries or electronic states is usually orders of magnitude smaller,
meaning that qualitatively correct estimates can typically be achieved
with just a few basis functions per electron.

Due to the systematic truncation error, LCAO basis sets often allow
good compromises between computational speed and accuracy, as basis
sets of various sizes and accuracies are available. Single-$\zeta$
basis sets may be used for preliminary studies where a qualitative
level of accuracy suffices; for instance, for determining the occupation
numbers in different orbital symmetry blocks, or the spin multiplicity
of the ground state of a well-behaved molecule. Polarized double-
or triple-$\zeta$ basis sets are already useful for quantitative
studies within DFT: thanks to the exponential basis set convergence
of SCF approaches and to the fortunate cancellation of errors mentioned
above, these basis sets often afford results that are sufficiently
close to the CBS limit. Quadruple-$\zeta$ and higher basis sets are
typically employed for benchmark DFT calculations, for the calculation
of molecular properties that are more challenging to describe than
the relative energy, as well as for post-HF calculations which require
large basis sets to reproduce reliable results.

Again allowed by the small number of basis functions, post-HF methods
such as Møller--Plesset perturbation theory\citep{Moller1934}, configuration
interaction (CI) theory and coupled-cluster (CC) theory\citep{Cizek1966}
can often be applied in the full space of molecular orbitals in LCAO
calculations. Because of this, the use of post-HF methods becomes
unambiguous, allowing the definition of trustworthy model chemistries\citep{Pople1999}
and enabling their routine applications to systems of chemical interest.

Finally, once again facilitated by the small number of basis functions,
several numerically robust methods are tractable for the solution
of the SCF equations of \eqref{Roothaan} within LCAO approaches.
As the direct iterative solution of \eqref{Roothaan} typically leads
to oscillatory divergence, the solution of the SCF problem is traditionally
stabilized with techniques such as damping,\citep{Cances2000} level
shifts,\citep{Saunders1973,Mitin1988,Domotor1989} or Fock matrix
extrapolation\citep{Kudin2002,Hu2010} in the initial iterations.
Once the procedure starts to approach convergence, the solution can
be accelerated with \emph{e.g.} the direct inversion in the iterative
subspace (DIIS) method,\citep{Pulay1980,Pulay1982} or approached
with true\citep{Wessel1967,Douady1979,Douady1980,Bacskay1981,Bacskay1982,HeadGordon1989,Chaban1997,Francisco2004,Thogersen2004,Thogersen2005}
or approximate\citep{Neese2000,VanVoorhis2002,Dunietz2002,Yang2007a}
second-order orbital rotation techniques that converge even in cases
where DIIS methods oscillate between solutions. For more information
on recently suggested Fock matrix extrapolation techniques, I refer
to \citerefs{Kudin2007, Garza2012, Garza2015a} and references therein.

\subsection{Variants of LCAO\label{subsec:Variants-of-LCAO}}

Three types of LCAO basis sets are commonly used. First, consider
the Slater-type orbitals\citep{Zener1930,Slater1930,Eckart1930} (STOs),
which are also commonly known as exponential type orbitals (ETOs)
\begin{equation}
R_{nl}(r)\propto r^{n-1}\exp(-\zeta_{nl}r).\label{eq:STO}
\end{equation}
STOs were especially popular in the early days of quantum chemistry,
since they represent analytical solutions to the hydrogenic Schrödinger
equation. The STO atomic wave functions tabulated by Clementi and
Roetti\citep{Clementi1974} should especially be mentioned here due
to their wealth of applications. For instance, their wave functions
are still sometimes used for non-self-consistent benchmarks of density
functionals.\citep{Chakraborty2011,Laricchia2014,Sun2015}

STOs have the correct asymptotic form both at the nucleus -- the
Kato cusp condition\citep{Kato1957} can be satistied -- as well
as far away, where the orbitals should exhibit exponential decay.\citep{Katriel1980,Silverstone1981,Mayer2003}
However, STOs are tricky for integral evaluation in molecular calculations,
typically forcing one to resort to numerical quadrature.\citep{Boerrigter1988,TeVelde1991,TeVelde2001,Franchini2013}
 Note that even though STO basis sets with non-integer $n$ values\citep{Parr1957,Joy1958,Saturno1958,Saturno1960}
have been shown to yield significantly more accurate absolute energies
than analogous basis sets that employ integer values for $n$,\citep{Koga1997,Koga1997a,Koga1997b,Koga1998,Koga2000a,Guseinov2009a}
they have not become widely used. Closely related to STOs, Coulomb
Sturmians\citealp{Shull1959,Rotenberg1962} have recently gained renewed
attention but have not yet become mainstream, see \emph{e.g.} \citerefs{Avery2012}
and \citenum{Herbst2018} and references therein.

Second, Gaussian-type orbitals (GTOs)
\begin{equation}
R_{nl}(r)\propto r^{l}\exp(-\alpha_{nl}r^{2})\label{eq:GTO}
\end{equation}
enjoy overwhelming popularity thanks to the Gaussian product theorem\citep{Boys1950}
that facilitates the computation of matrix elements, which is especially
beneficial for the two-electron integrals,\citep{McMurchie1978,Obara1986,Gill1991}
the calculation of which is typically the rate determining step in
SCF calculations. The importance of efficient, analytical evaluation
of molecular integrals can hardly be overstated. Efficient integral
evaluation makes calculations affordable: it has long been possible
to use large atomic basis sets and/or study large systems with Gaussian
basis sets. Analytical integral evaluation guarantees accuracy: it
is straightforward to calculate \emph{e.g.} geometric derivatives
and reliably estimate force constants and vibrational frequencies,
because the matrix elements are numerically stable.

Another reason for the popularity of GTOs was the recognition early
on that STOs can in principle be expanded exactly in GTOs;\citep{Kikuchi1954}
thus, for instance, the hydrogenic atom can be solved with GTO expansions
with controllable accuracy, highlighting the flexibility of Gaussian
basis sets.\citep{Hehre1969} Indeed, most commonly-used quantum chemistry
programs, such as \citerefs{Schmidt1993, Guest2005, Werner2012, Neese2012, Aidas2014, Shao2015, Gaussian16, Aquilante2016, Dirac2017, Parrish2017, Sun2018}
to name a few, employ Gaussian basis sets such as the Pople 6-31G\citep{Ditchfield1971,Hehre1972,Hariharan1973,Hariharan1974,Gordon1980,Francl1982,Binning1990,Blaudeau1997,Rassolov1998,Rassolov2001a}
and 6-311G series,\citep{McLean1980,Krishnan1980} the Karlsruhe def\citep{Schafer1992,Schafer1994}
and def2 series,\citealp{Weigend2005,Rappoport2010} the correlation
consistent cc-pVXZ series by Dunning, Peterson, and coworkers,\citealp{Dunning1989,Kendall1992,Woon1993,Woon1994,Woon1995,Wilson1996,Prascher2011,VanMourik2000,Wilson1999,Dunning2001,Peterson2002,Hill2017}
the polarization consistent pc-$n$ and pcseg-$n$ series by Jensen,\citealp{Jensen2001,Jensen2002,Jensen2002b,Jensen2007,Jensen2004,Jensen2012,Jensen2013,Jensen2014}
the $n$ZaPa sets by Petersson and coworkers,\citep{Ranasinghe2013,Ranasinghe2015}
as well as the atomic natural orbital (ANO) basis sets by Almlöf,
Roos, and coworkers;\citep{Almlof1987,Almlof1988,Almlof1990,Widmark1990,Widmark1991,Pierloot1995,Pou-Amerigo1995,Roos2004,Roos2004a,Roos2005,Roos2005a,Roos2008}
see \emph{e.g.} \citerefs{Peterson2011a, Hill2013, Jensen2013b} for
reviews on Gaussian basis sets.

Third, numerical atomic orbitals (NAOs)
\begin{equation}
R_{nl}(r)=\frac{u_{nl}(r)}{r}\label{eq:NAO}
\end{equation}
are the most recent addition to the field.\citep{Averill1973,Delley1982,Delley1990,Talman1993}
Unlike GTOs and STOs, NAOs have no analytical form; instead, NAOs
are typically pretabulated on a radial grid. (Note that adaptive approaches
employing flexible NAOs have also been suggested but have not become
commonly used;\citep{Talman2000,Andrae2001,Talman2003,Talman2010,Talman2010a}
the atomic radial grid approach should also be mentioned in this context.\citep{Becke1990a,Shiozaki2007})

The NAO form is \emph{by definition} superior to GTOs or STOs: a suitably
parametrized NAO basis set is able to represent GTOs and STOs exactly,
while the opposite does not hold. Thanks to this great flexibility,
unlike GTOs or STOs, a minimal NAO basis set can be an \emph{exact}
solution to the noninteracting atom at the SCF level of theory: a
single NAO can simultaneously satisfy both the Kato cusp condition
at the nucleus \emph{and} have the correct exponential asymptotic
behavior.

In further contrast to primitive \emph{i.e.} uncontracted STO and
GTO basis sets, single-center NAOs can be defined at no additional
cost to be orthonormal, which leads to better conditioned equations
even for the molecular case. These features make NAOs the most accurate
LCAO basis set, as has been recently demonstrated in multiple studies
both at the SCF and post-HF levels of theory.\citep{Blum2009,Zhang2013,Lejaeghere2016,Jensen2017}
Alike STOs, NAOs require quadrature for molecular integral evaluation,
but this does not appear to be a problem on present-day computers.\citep{Delley2000,Blum2009}
NAOs are typically constructed from SCF calculations on atoms, which
will be reviewed in the next section.

\subsection{Case study: Kr atom; contracted basis sets}

\begin{figure}
\begin{centering}
\subfloat[$1s$]{\begin{centering}
\includegraphics[width=0.33\textwidth]{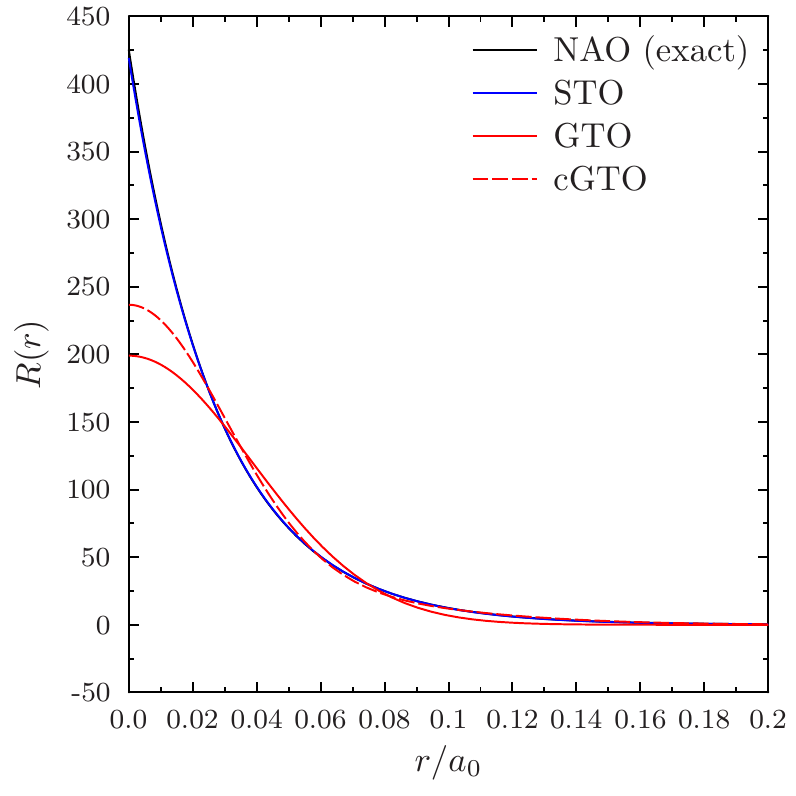}
\par\end{centering}
}\subfloat[$2p$]{\begin{centering}
\includegraphics[width=0.33\textwidth]{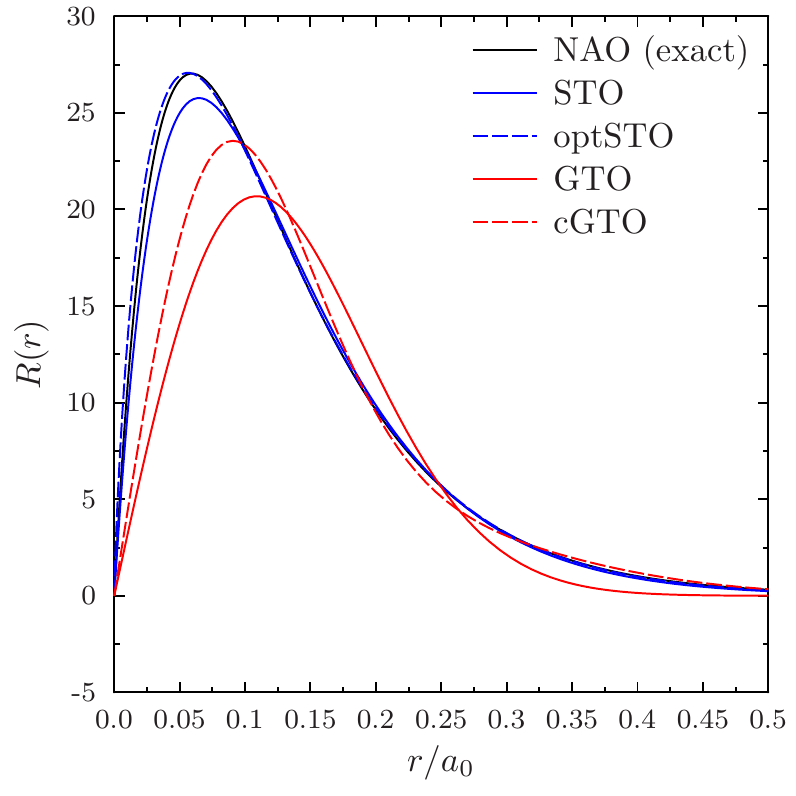}
\par\end{centering}

}\subfloat[$3d$]{\begin{centering}
\includegraphics[width=0.33\textwidth]{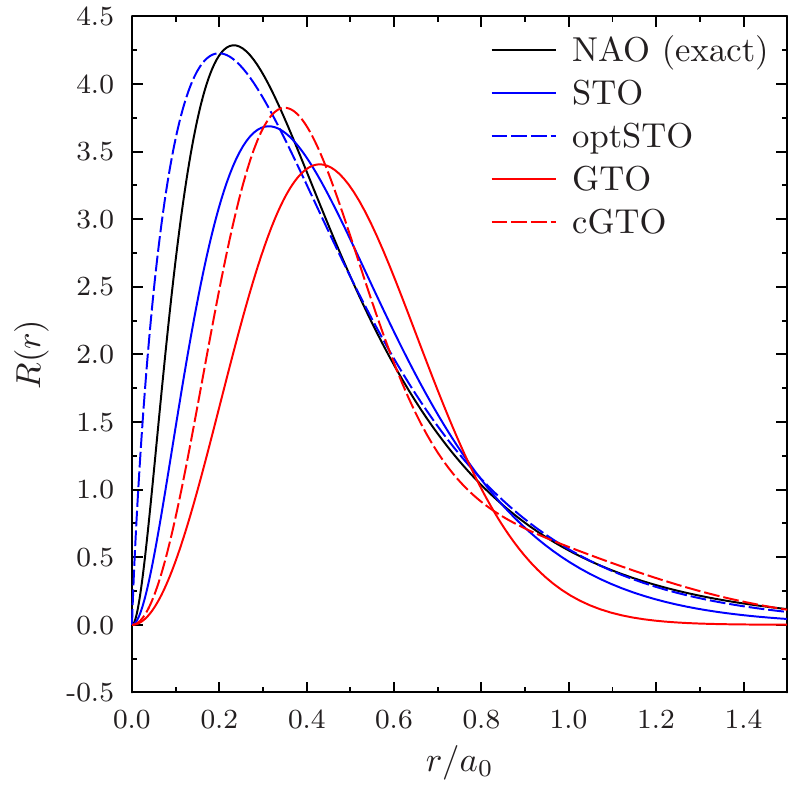}
\par\end{centering}
}
\par\end{centering}
\caption{Comparison of NAOs, STOs, GTO, and contracted GTOs (cGTOs) for the
$1s$, $2p$, and $3d$ radial functions of krypton. STOs with an
optimized fractional value for $n$ are also shown (optSTO).\label{fig:basis-comparison}}
\end{figure}

A comparison of the three types of basis functions for an exchange-only
local density calculation on the krypton atom is shown in \figref{basis-comparison}
for the atomic $1s$, $2p$, and $3d$ orbitals. Here, the NAOs are
chosen as the exact $1s$, $2p$, and $3d$ orbitals, whereas the
STO and GTO functions have been optimized for best overlap with the
exact orbital.

The STO form clearly allows an accurate expansion of the $1s$ orbital.
The $1s$, however, is a special case, as it experiences the least
screening by the other electrons. These screening effects deform the
orbital from the hydrogenic STO form, and become more visible for
the $2p$ and $3d$ orbitals: for $3d$, the STO orbital is already
quite dissimilar from the exact one. But, as was discussed above in
\subsecref{Variants-of-LCAO}, relaxing the restrictions in the STO
form from $n=2$ and $n=3$ for the $2p$ and $3d$ orbitals, respectively,
to also allow fractional $n$ leads to significant improvements in
the accuracy of the STO orbitals, as can be seen in \figref{basis-comparison}.

There are large differences between the GTO and the exact orbitals,
which are especially clear for the $1s$ orbital which should have
a cusp at the origin. However, the accuracy of the GTO basis set can
be radically improved by adding more functions. \Figref{basis-comparison}
also shows results for contracted GTO (cGTO) basis functions, which
have been formed from two GTO primitive functions by optimizing the
primitives and their expansion coefficients for the best overlap with
the exact orbital. The cGTO basis set can be made more and more accurate
by adding more and more primitives to the expansion.

However, it is worth pointing out here that in a sense, cGTOs are
also a form of a NAO basis. Employing general contractions,\citep{Raffenetti1973}
the single-center basis functions can be chosen to be strictly orthonormal,
and non-interacting atoms can be solved to high accuracy in a minimal
basis set that is assembled from a large number of primitive Gaussians.
Alike NAOs, efficient integral evaluation in generally contracted
GTO (gcGTO) basis sets is significantly more complicated than in primitive
or segmented Gaussian basis sets, since a primitive integral can contribute
to several contracted integrals.

gcGTO basis sets are typically formed from ANOs, and they are well-known
to yield faster convergence to the basis set limit with respect to
the number of basis functions than uncontracted basis sets or basis
sets employing segmented contractions.\citep{Almlof1987,Almlof1988,Almlof1990,Illas1990,Widmark1990,Widmark1991,Bauschlicher1993,Pierloot1995,Pou-Amerigo1995,Roos2004,Roos2004a,Roos2005,Roos2005a,Roos2008,Neese2011}
Even though gcGTOs are similar to NAOs, the number of degrees of freedom
in any Gaussian basis set expansion is much smaller than what is possible
with a true NAO basis set: large Gaussian basis sets for heavy atoms
typically contain up to 30 or 40 primitives per angular momentum shell,
whereas NAOs are often tabulated with hundreds to thousands of radial
points. As NAOs allow for a more accurate description of atomic orbitals
than Gaussian expansions do, while ANOs yield faster convergence in
post-HF calculations than basis sets constructed from HF orbitals
do, it is clear that it should be possible to form extremely efficient
numerical ANO (NANO) basis sets for post-HF studies. The generation
of NANO basis sets could follow pre-established procedures to generate
ANOs from Gaussian expansions; suitable fully numerical implementations
of the multiconfigurational approaches that are typically used to
construct ANOs have been described in the literature, as is discussed
below in \subsecref{Atoms}.

\subsection{Deficiencies of LCAO approaches\label{subsec:Deficiencies-of-LCAO}}

In the case of atomic calculations, it is relatively straightforward
to obtain near-CBS limit energies even with LCAO basis sets: since
only a single expansion center is involved, the basis set is relatively
well-conditioned even if large basis set expansions are used. However,
the basis set requirements depend not only on the level of theory
and property; they may also depend on the studied state.

For some properties, it is possible to explicitly identify the regions
of the basis set that require extra focus: for instance, the Fermi
contact term for a nucleus A is proportional to $\propto\delta(\boldsymbol{r}_{A})$,
where $\boldsymbol{r}_{A}$ is the distance from nucleus, so an accurate
reproduction of the electronic structure at the nucleus is necessary.
Because only $s$ orbitals are non-zero at the nucleus, the accuracy
of the description of the contact term is mostly dependent on the
tight $s$ functions.

Other examples include the polarization induced by weak electric and/or
magnetic field perturbations, for which analytic expressions can be
derived for the change in the atomic orbitals. Equipped with this
knowledge, one can parametrize perturbation-including basis sets both
in the case of STOs\citep{Zeiss1979,Chong2003,Rossikhin2014} and
GTOs\citep{Sadlej1977,Wolinski1979,Roos1985,Wolinski1985,Sadlej1988,Sadlej1991,Sadlej1991a,Sadlej1991b,Kello1996,Miadokova1997,Cernusak2003,Chong2003,Baranowska2007,Baranowska2009,Baranowska2009a,Voronkov2012}
which achieve faster convergence to the basis set limit than the non-perturbation-including
basis sets. Similarly, knowledge on the asymptotic long-range behavior
of the wave function\citep{Silverstone1981,Mayer2003} can be used
to construct GTO basis sets with improved asymptotic behavior.\citep{Koga1992}

Completeness-optimization\citep{Manninen2006,Lehtola2015} is a general
approach that may be used to parametrize basis sets that yield near-CBS
values for the energy and/or whatever molecular properties. The approach
does not make any assumptions on the level of theory or specific properties
that are calculated; instead, the only assumption is that the property
can be converged with a sufficiently large basis set. The approach
relies on an iterative approach where new basis functions are added
one by one, until the property has converged. Completeness-optimization
has been demonstrated so far for electric and magnetic properties,\citep{Manninen2006,Ikalainen2008,Ikalainen2009,Ikalainen2010,Lantto2011,Ikalainen2012,Fu2013,Vaara2013,Abuzaid2013,Vahakangas2013,Vahakangas2014,Hanni2017,Lehtola2015}
electron momentum densities,\citep{Lehtola2012a,Lehtola2013} as well
as NAO simulations of nanoplasmonics.\citep{Rossi2015}

However, getting to the CBS limit can still be painstaking, as several
sets of tight and/or diffuse functions may be necessary to achieve
converged results.\citep{Sim1992,Woon1994,Helgaker1998,Jensen2006,Jensen2008}
For instance, Gaussian primitives with exponents as small as $6.9\times10^{-9}$
have been found necessary to make \ce{F-} properly bound at the DFT
level.\citep{Jarecki1999} Basis sets with a fixed analytic form also
pose intrinsic limits to their accuracy because of their incorrect
asymptotic behavior. This has been shown in the case of Gaussian basis
sets \emph{e.g.} for the moments of the electron momentum density,\citep{Simas1982,Lehtola2012a,Lehtola2013}
the electron density,\citep{Christiansen1978a,Simas1983,Mastalerz2008,Mastalerz2010}
as well as nuclear magnetic properties.\citep{Jensen2016} Reaching
the CBS limit for such properties may be untractable unless more flexible
basis sets are employed.

Molecular calculations are even more difficult than atomic ones. Although
it is straightforward to establish a hierarchy between NAO, STO, and
GTO basis sets in the case of atoms and optimize basis sets for them,
atomic symmetries are broken by the molecular environment, requiring
the addition of \emph{polarization functions} into the atomic basis
set. It is generally acknowledged that the optimal form of polarization
functions is not known, and that the ranking of GTOs \emph{vs }STOs
\emph{vs }NAOs is not clear in the case of polarization functions.
Indeed, GTO and STO polarization functions are commonly employed in
NAO calculations.\citep{Larsen2009,Blum2009}

Another complication is that the set of AO basis functions placed
on the various nuclei may develop linear dependencies. Linear dependencies
are especially problematic for diffuse basis functions, whereas tight
functions are less problematic since they are highly localized. Thus,
it is especially difficult to accurately reproduce molecular properties
that are sensitive to the diffuse part of the wave function with LCAO
approaches; further complications may also be caused by the limitations
in the form of the AO basis functions discussed above.\citep{Christiansen1978a,Simas1982,Simas1983,Mastalerz2008,Mastalerz2010,Lehtola2012a,Lehtola2013,Jensen2016}

Any linear dependencies that exist in the basis set are typically
removed via the canonical orthonormalization procedure\citep{Lowdin1956}
in order to make the basis unambiguous. In such a case, a calculation
with a smaller basis set can reproduce the same or a lower energy
than the one predicted by the calculation with the linearly dependent
basis.

There are workarounds to this problem for some cases. Instead of using
a large basis set on each of the nuclei, one can employ bond functions:
use a smaller basis on the nuclei while placing new basis functions
in the middle of chemical bonds; such a procedure can yield faster
convergence to the basis set limit.\citep{Matito2006a} Moreover,
one can replace diffuse functions placed on every atom with ones placed
only on the molecular center of symmetry. However, in either case,
the optimal division and placement of functions to recover the wanted
property yet keep linear dependencies at bay may be highly nontrivial
for a general system, and is clearly dependent on the molecular geometry.

A general approach for avoiding the problems with linear dependencies
that arise from diffuse functions in a molecular setting has, however,
been suggested. In this approach, diffuse functions are modeled by
a large number of non-diffuse functions placed all around the system.\citep{Melichercik2013,Melichercik2018}
The use of such lattices of Gaussian functions is an old idea, originally
proposed as the ``Gaussian cell model'' by Murrell and coworkers.\citep{Haines1974}
The Gaussian cell model was later revisited by Wilson and coworkers,
who showed that much better results can be obtained by supplementing
the lattice basis set with AO functions on the nuclei.\citep{Ralston1995,Wilson1995a,Wilson1996a}
The authors of \citerefs{Melichercik2013} and \citenum{Melichercik2018}
appear to have been unaware of the work by Murrell, Wilson, and coworkers.
Nonetheless, the Gaussian cell model is but an imitation of proper
real-space methods: the use of a grid of spherically symmetric Gaussians
clearly cannot yield the same amount of accuracy or convenience as
a true real-space approach that employs a set of systematically convergent
basis functions with finite support. In one dimension, however, arrays
of Gaussian functions called gausslets have been recently shown to
yield useful completeness properties.\citealp{White2017} But, it
is not yet clear how well the gausslet approach can be generalized
to atoms heavier than hydrogen, or to general three-dimensional molecular
structures.

The accurate calculation of \emph{e.g. }potential energy surfaces
$V_{AB}=E(A\cdots B)-E(A)-E(B)$ is also complicated in the LCAO approach
due to the so-called basis set superposition error (BSSE). $E(A\cdots B)$
can be underestimated with respect to $E(A)$ and $E(B)$, because
the locality of the electrons onto the basis functions belonging to
the fragments is not enforced. Although BSSE can be approximately
removed with the Jansen--Ros--Boys--Bernardi counterpoise correction\citep{Jansen1969,Boys1970}
or developments thereof,\citep{Wells1983,Martin1989,White1990,Turi1993,Valiron1997,Mierzwicki2003}
and although BSSE becomes smaller when going to larger atomic basis
sets, it is not always clear how accurate the obtained interaction
energies are. The geometry-dependence of the molecular basis set also
complicates the calculation of molecular geometries and vibrational
frequencies, as Pulay terms\citep{Pulay1969} appear.

Finally, it is difficult to improve the description of the wave function
in an isolated region of space in the LCAO approach. The addition
of a new function to describe \emph{e.g. }the wave function close
to the nucleus requires changing the solution \emph{globally} to maintain
orbital orthonormality, because the basis functions are non-orthogonal
and have global support. Could these problems be circumvented somehow?

\section{Real-space methods\label{sec:Real-space-methods}}

\begin{figure}
\begin{centering}
\includegraphics[width=0.5\textwidth]{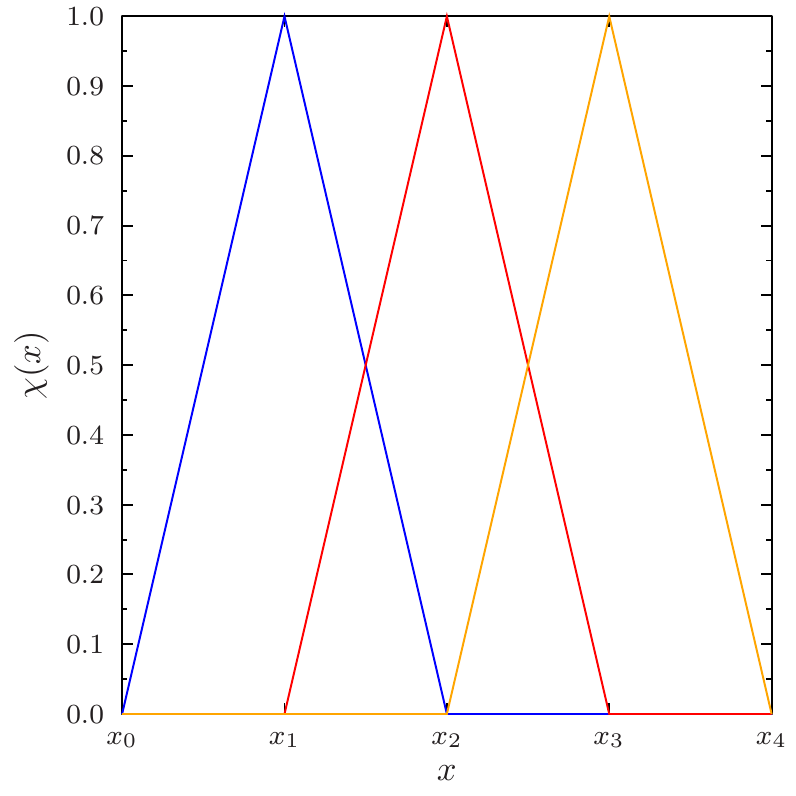}
\par\end{centering}
\caption{Illustration of hat functions, which are non-zero only within a finite
interval.\label{fig:Hat-functions.}}

\end{figure}

An alternative to the use of LCAO basis sets is to switch to real-space
methods, where the basis functions no longer have a form motivated
purely by chemistry. Instead of expanding the orbitals in terms of
AOs as in \eqref{LCAO}, the expansion is generalized to
\begin{equation}
\psi_{i}^{\sigma}(\boldsymbol{r})=\sum_{I}C_{Ii}^{\sigma}\Xi_{I}(\boldsymbol{r}),\label{eq:realspace}
\end{equation}
where the real-space basis set $\Xi_{I}(\boldsymbol{r})$ can be chosen
based on convenience for the problem at hand. For illustration, a
simple option is to form the three-dimensional basis set $\Xi_{I}(\boldsymbol{r})$
as a tensor product of one-dimensional basis functions $\chi_{i}(r)$
as

\begin{equation}
\Xi_{I}(\boldsymbol{r})=\chi_{I_{x}}(x)\chi_{I_{y}}(y)\chi_{I_{z}}(z),\label{eq:tensorbasis}
\end{equation}
where the $\chi_{i}(r)$ can in turn be chosen as \emph{e.g. }the
hat functions shown in \figref{Hat-functions.}. In many dimensions,
also other kinds of choices for the three-dimensional basis functions
can be made instead of the rectangular elements of \eqref{tensorbasis}:
for instance, triangles in two dimensions, or tetrahedra, hexahedra,
or prisms in three dimensions; more information on these can be found
in any standard finite element textbook such as \citeref{Zienkiewicz2013}.

The present work focuses on fully numerical calculations on atoms
and diatomic molecules. As will be reviewed below in \secref{Applications},
a one-dimensional treatment is sufficient for atoms, whereas diatomic
molecules require a two-dimensional solution. Almost all the diatomic
approaches reviewed below use a rectangular basis similar to \eqref{tensorbasis}:
\begin{equation}
\Xi_{I}(\xi,\eta)=\chi_{I_{\xi}}(\xi)\chi_{I_{\eta}}(\eta),\label{eq:2Dbasis}
\end{equation}
where $\xi$ and $\eta$ are prolate spheroidal coordinates introduced
below in \subsecref{Diatomic-molecules}. The only exception is the
program by Heinemann, Fricke and Kolb which employs triangular elements
as discussed in \citeref{Heinemann1988a}; the present discussion
can thus be restricted to the case of \eqref{tensorbasis} or \eqref{2Dbasis}.
Now, as the basis factorizes in the coordinates, it is sufficient
to restrict the discussion to one dimension, as the generalization
to many dimensions is obvious.

Four kinds of numerical approaches are commonly used for the one-dimensional
basis functions $\chi_{I}(x)$: finite differences, finite elements,
basis splines, and the discrete variable representation. The common
feature in all the approaches is that a function $f(x)$ is represented
in terms of the basis functions $\chi_{I}(x)$ as 
\begin{equation}
f(x)\approx\sum_{I}c_{I}\chi_{I}(x).\label{eq:fappr}
\end{equation}
As will be seen below, the first three methods rely on a local piecewise
polynomial representation for $f(x)$, whereas the last one can also
be formulated in terms of non-polynomial basis functions $\chi_{I}(x)$.
All approaches should yield the same value at the CBS limit -- this
is what makes fully numerical approaches enticing in the first place.

However, as will be discussed below, the speed of convergence to the
CBS limit depends heavily on the order of the used numerical approximation,
with higher-order methods affording faster convergence in the number
of basis functions than lower-order methods. This means that comparing
the accuracy of the various approaches is not trivial, as even the
same approach can manifest different convergence properties when implemented
at different orders.

Furthermore, the speed of convergence also depends in each case on
the system, the studied property, and the level of theory used in
the calculation: HF and DFT afford exponential converge in the basis
set, whereas post-HF methods only exhibit polynomial convergence if
the orbital space is not fixed. For this reason, for numerical results
I refer the reader to the individual references listed in \secref{Applications}.

\subsection{Finite differences\label{subsec:Finite-differences}}

The finite difference method is arguably the oldest and the best-known
method for numerically solving differential equations. The method
is based on a Taylor expansion of $f(x)$ as
\begin{equation}
f(x\pm h)=f(x)\pm f'(x)h+\frac{1}{2}f''(x)h^{2}+\mathcal{O}(h^{3}),\label{eq:taylor}
\end{equation}
where it is assumed that $f(x)$ is tabulated on a grid $x_{i}=x_{0}+ih$
as $f_{i}=f(x_{i})$, where $h$ is the grid spacing; that is, $f(x_{i}+h)=f_{i+1}$.
From the Taylor expansion, \eqref{taylor}, one can obtain approximations
for the derivatives of $f(x)$, for instance, using central differences
as
\begin{align}
f'(x) & \approx\frac{f(x+h)-f(x-h)}{2h}+\mathcal{O}(h^{2}),\label{eq:df-twop}\\
f''(x) & \approx\frac{f(x+h)-2f(x)+f(x-h)}{h^{2}}+\mathcal{O}(h^{2});\label{eq:d2f-threep}
\end{align}
these are known as the two-point and three-point stencils, respectively.
The procedure can also be done at higher orders, leading to \emph{e.g.
}the four- and five-point rules

\begin{align}
f'(x) & \approx\frac{-f(x+2h)+8f(x+h)-8f(x-h)+f(x-2h)}{12h}+\mathcal{O}(h^{4}),\label{eq:df-fourp}\\
f''(x) & \approx\frac{-f(x+2h)+16f(x+h)-30f(x)+16f(x-h)-f(x-2h)}{12h^{2}}+\mathcal{O}(h^{4}),\label{eq:d2f-fivep}
\end{align}
which carry a smaller error, as evidenced by the smaller remainder
term. Finite difference formulas can be easily generated to arbitrary
order by recursion, even on nonuniform grids.\citep{Fornberg1988}

An important feature of the finite difference method is that relaxation
approaches can be formulated for Schrödinger-type equations. For instance,
applying \eqref{df-twop} onto
\begin{equation}
-\frac{d^{2}\psi}{dx^{2}}(x)+V(x)\psi(x)=E\psi(x)\label{eq:fd-schr}
\end{equation}
one obtains
\begin{equation}
-\frac{\psi_{i+1}-2\psi_{i}+\psi_{i-1}}{h^{2}}+V_{i}\psi_{i}=E\psi_{i}\label{eq:fd-schr-threep}
\end{equation}
which can be rearranged into the relaxation equation
\begin{equation}
\overline{\psi}_{i}=\frac{\psi_{i-1}+\psi_{i+1}}{2}+\frac{1}{2}\left(V_{i}-E\right)\psi_{i}h^{2}\label{eq:fi-relax}
\end{equation}
that can be solved by iteration. Furthermore, the iteration $\psi_{i}^{(0)}\to\psi_{i}^{(1)}\to\dots\to\psi_{i}^{(k)}\to\dots$
can be made faster by the use of the successive overrelaxation method
\begin{equation}
\psi_{i}^{(k)}=\omega\overline{\psi}_{i}^{(k)}+(1-\omega)\psi_{i}^{(k-1)}\label{eq:SOR}
\end{equation}
where $0<\omega\leq1$ corresponds to relaxation and $\omega>1$ to
overrelaxation.

Although the relaxation approach is straightforward and readily applicable
to high-performance computing, the drawback is that a good initial
guess is necessary for the solution $\psi_{i}$ as well as the eigenvalue
$E$. Another drawback is that, as can be seen from \eqref{fi-relax},
information is passed only locally, and so the relaxation becomes
extremely slow in fine meshes; this can be, however, circumvented
with multigrid approaches\citep{Hackbusch1985} that pass information
simultaneously at multiple length scales, thus avoiding the slowdown.

It is also possible to avoid using relaxation approaches altogether:
finite difference calculations can be written as matrix eigenvalue
problems, e.g. \eqref{fd-schr-threep} can be rewritten as
\begin{equation}
\boldsymbol{H}\boldsymbol{\psi}=\boldsymbol{E}\boldsymbol{\psi}\label{eq:fd-mat}
\end{equation}
and the finite-difference Hamiltonian is given by the banded-diagonal
matrix
\begin{equation}
\boldsymbol{H}=\left(\begin{array}{ccccccc}
\ddots & \ddots\\
\ddots & -h^{-2} & 2h^{-2}+V_{i-1} & -h^{-2}\\
 &  & -h^{-2} & 2h^{-2}+V_{i} & -h^{-2}\\
 &  &  & -h^{-2} & 2h^{-2}+V_{i+1} & -h^{-2} & \ddots\\
 &  &  &  &  & \ddots & \ddots
\end{array}\right),\label{eq:Hfd}
\end{equation}
where $\boldsymbol{\psi}$ and $\boldsymbol{V}$ are vectors of the
values of the wave function and the potential, respectively; however,
this is rarely done in practice, as the basis spline and finite element
methods described below naturally lead to matrix equations.

Having obtained approximate derivatives of the function in the vicinity
of the current point using \emph{e.g. }\eqref{df-twop, d2f-threep}
or \eqref{df-fourp, d2f-fivep}, \eqref{taylor} expands the description
of the function beyond the known values $f_{i}$ to how the function
behaves in-between the grid points. From the form of \eqref{taylor},
it is clear that the finite difference method in fact corresponds
to a local, piecewise polynomial basis for $f(x)$. However, the Taylor
rule of \eqref{taylor} implies that a different polynomial is used
for every grid point $x_{i}$, which leads to a discontinuous representation
for $f(x)$. This is the puzzle of the finite difference method: it
is not obvious how the approximated function behaves in-between grid
points.

Applications to quantum mechanics require matrix elements that involve
spatial integrals; thus, a quadrature rule is necessary, but due to
the aforementioned ambiguity, it is not possible to derive one from
the finite difference formulation. Instead, the typical solution is
to just use quadrature formulas such as Newton--Cotes, which can
be derived from Lagrange interpolating polynomials, yielding \emph{e.g.}
the two-point trapezoid rule, the three-point Simpson's rule, and
the five-point Boole's rule. Although these quadrature formulas imply
using the same fitting polynomial for all points $\{x_{i}\}$ in the
fitting integral -- contrasted to using \emph{different} polynomials
at every point to get the derivatives according to \eqrangeref{df-twop}{d2f-fivep}
-- there \emph{is} a solid mathematical argument for using such a
hybrid scheme:\emph{ }the error bounds. For instance, the trapezoid
and Simpson's rules have errors of $\mathcal{O}(h^{3})$ and $\mathcal{O}(h^{5})$,
respectively, which can be contrasted with the $\mathcal{O}(h^{2})$
and $\mathcal{O}(h^{4})$ errors for the derivatives in \eqref{df-twop, d2f-threep}
or \eqref{df-fourp, d2f-fivep}, respectively. Because of this, the
mismatch between the rules for derivation and integration is not a
problem in practice: numerical integration is more accurate than numerical
differentiation, and an arbitrary level of accuracy is anyhow achievable
by using a finer grid $h\to0$.

Still, getting to the CBS limit is complicated by the lack of variationality
in finite difference calculations:\citep{Beck2000} when the variational
principle does not hold, calculations with a smaller $h$ can yield
an \emph{increased} energy, even though a better basis set is used.
This error is caused by the underestimation of the kinetic energy
by the finite difference representation of the Laplace operator.\citep{Maragakis2001,Skylaris2002a}

But, finite difference electronic structure calculations often have
also another, even larger source of non-variationality than the kinetic
energy approximation: aliasing error in the Coulomb and exchange potentials.
This error arises whenever the electron density is expressed in the
same grid as the molecular orbitals
\begin{equation}
\rho_{n}=\rho\left(\boldsymbol{r}_{n}\right)=\sum_{i\text{ occupied}}\left|\psi_{i}(\boldsymbol{r}_{n})\right|^{2}=\sum_{i\text{ occupied}}\left|\psi_{ni}\right|^{2}\label{eq:aliasing}
\end{equation}
because an order-$k$ polynomial wave function yields a density that
has an order-$2k$ polynomial expansion, which ends up being truncated.
The aliasing error is analogous to the error of resolution-of-the-identity
approaches of LCAO methods:\citep{Baerends1973,Sambe1975,Mintmire1982,Vahtras1993,Feyereisen1993}
it makes the classical Coulomb interaction between the electrons less
repulsive, and the exchange interaction less attractive. The latter
error yields a slight increase of the energy, but the former error
makes the energy too negative, resulting in energies that are underestimated
i.e. anti-variational until the grid is made fine enough.

One more limitation in the finite difference method is that if an
equispaced grid is used, higher-order rules which are based on high-order
polynomials become unstable due to Runge's phenomenon.\citep{Runge1901}
For this reason, for instance the \textsc{x2dhf} program\citep{Kobus1996,Kobus2013,x2dhf}
discussed below in \subsecref{Diatomic-molecules} employs eight-order
\emph{i.e. }9-point central difference expressions for computing second
derivatives with $\mathcal{O}(h^{8})$ error, and a 7-point rule for
quadrature with $\mathcal{O}(h^{9})$ error.\citep{Kobus2013}

\subsection{Explicit basis set methods\label{subsec:Explicit-basis-set}}

Instead of invoking a local piecewise polynomial basis only implicitly
like in the finite difference method, the finite element method (FEM)
explicitly uses a piecewise polynomial basis set to represent the
function $f(x)$ like in interpolation theory. Because of this, it
can be remarked that finite differences approximate the Schrödinger
equation, while finite elements -- as well as basis splines and the
discrete variable representation -- approximate its solution. The
nice thing about these three approaches is that the basis is fully
defined: evaluating the function $f(x)$ or its derivatives $f^{(n)}(x)$
at an arbitrary point $x$ is completely unambiguous, meaning \emph{e.g.}
that adaptive quadrature rules, or rules relying on points other than
the discretization nodes can be easily employed. Moreover, because
an unambiguous basis set has now been defined, instead of relying
on an (over)relaxation approach as in the finite difference method,
one may use the same variational techniques as in LCAO approaches
to solve for the wave functions, \emph{e.g.} the Roothaan--Hall equation
of \eqref{Roothaan}.

In the finite element method, the domain for $x$ is divided into
into line segments called elements $x\in[x^{a},x^{b}]$, and a separate,
local polynomial basis is defined within each element. The basis functions
within an element, commonly known also as \emph{shape functions},
are determined by choosing $n$ points $x_{i}$ called \emph{nodes}
within the element, $x_{i}\in[x^{a},x^{b}]$.

In the most commonly used finite element variant, one demands that
the value of $f(x)$ at each node is given by a specific basis function
\begin{equation}
f(x_{i})=\sum_{j}c_{j}\chi_{j}(x_{i})=f_{i}.\label{eq:fem}
\end{equation}
Employing a polynomial expansion for $\chi_{i}$, one sees that the
shape functions for $n$ nodes are $n-1$ order polynomials. Especially,
the condition $\chi_{j}(x_{i})=\delta_{ij}$ in \eqref{fem} is fulfilled
by Lagrange interpolating polynomials (LIPs)
\begin{equation}
\chi_{i}(x)=\prod_{j=0,j\neq i}^{n-1}\frac{x-x_{j}}{x_{i}-x_{j}},\label{eq:LIP}
\end{equation}
which are well-known to yield the error estimate
\begin{equation}
\left|f(x)-\tilde{f}(x)\right|\leq\frac{\prod_{i=0}^{n-1}(x-x_{i})}{n!}\max_{\xi\in[x^{a},x^{b}]}\left|f^{(n)}(\xi)\right|.\label{eq:LIP-error}
\end{equation}

Nodes are typically placed also at the element boundaries. Because
the node at the element boundary is shared by the rightmost function
in the element on the left, and the leftmost function in the element
on the right, these two basis functions have to be identified with
each other, guaranteeing continuity of $f(x)$ across the element
boundary. For instance, the hat functions in \figref{Hat-functions.}
correspond to two-node elements, where both nodes are on element boundaries.

Because all shape functions are fully localized within a single element,
it is very easy to take advantage of this locality in computer implementations.
Moreover, the correspondence of functions belonging to nodes at element
boundaries can be easily handled in computer implementations by index
manipulations, \emph{i.e.} overlaying.

In this formulation, because all basis functions are smooth, $f(x)$
will also be smooth within each element, and the function $f(x)$
is guaranteed to be continuous over element boundaries. However, while
all derivatives are guaranteed to be continuous within an element,
discontinuities in the first derivative $f'(x)$ may be seen at the
element boundaries.

Although the continuity of the derivative(s) can be strictly enforced
by straightforward extension of the approach in \eqref{fem}, in which
the description of the values of $f(x)$ and $f'(x)$ at the nodes
are each identified to arise from a single shape function, this does
not appear to be necessary in typical applications to quantum mechanics.
The reason for this is the variational principle: discontinuities
in the derivative imply a higher kinetic energy. The derivatives of
the Lagrange interpolating polynomials, \eqref{LIP}, do \emph{not}
vanish at the nodes, and this flexibility gives the variational principle
room to try to match the derivatives across element boundaries.

Although the Runge phenomenon\citep{Runge1901} is also a problem
for the finite element method with uniformly spaced nodes within the
element, the method can be made numerically stable to high orders
if one employs non-uniformly spaced nodes; suitable choices include
Gauss--Lobatto nodes, or Chebyshev nodes as in the spectral element
method.\citep{Patera1984} The benefit of high-order elements is apparent
in \ref{eq:LIP-error}: the errors become small rapidly in increasing
element order.\citep{Lehtola2019a} If in addition to defining the
shape functions, the same rule is also used for quadrature, one obtains
a discrete variable representation (DVR) which will be discussed in
more detail below.

In addition to varying the location of the nodes within each element
to make high-order rules numerically stable, it is possible to use
elements of different sizes and orders within a single calculation,
granting optimal convergence to the basis set limit with the least
possible number of basis functions; this is the origin for the popularity
of FEM across disciplines. For more discussion on the finite element
method, including other kinds of shape functions than Lagrange interpolating
polynomials, I refer to my recent discussion in \citeref{Lehtola2019a}.

Like the finite element method, basis splines (B-splines) are piecewise
polynomials, with an order $n$ spline corresponding to a polynomial
function of order $n$. B-splines were originally developed by Schoenberg,\citep{Schoenberg1946}
and were introduced in the context of atomic calculations by Gilbert
and Bentoncini,\citep{Gilbert1974,Gilbert1974a,Gilbert1975,Gilbert1975a}
based on earlier work on their application to single-particle problems
by Shore.\citep{Shore1973,Shore1973a,Shore1973b,Shore1974,Shore1975}
B-splines are defined by \emph{knots} $x_{0}\leq x_{1}\leq x_{n+1}$,
which need not be uniformly spaced, with the Cox--de Boor recursion
\begin{align}
B_{i,0}= & \begin{cases}
1 & x_{i}\leq x<x_{i+1}\\
0 & \text{otherwise}
\end{cases},\label{eq:B-prim}\\
B_{i,n}(x)= & \frac{x-x_{i}}{x_{i+n}-x_{i}}B_{i,n-1}(x)+\frac{x_{i+n+1}-x}{x_{i+n+1}-x_{i+1}}B_{i+1,n-1}(x);\label{eq:B-rec}
\end{align}
thus, $B_{i,0}$ are piecewise constants, $B_{i,1}$ are hat functions
(\figref{Hat-functions.}), and so on. On each interval $(x_{i},x_{i+1})$,
only $n$ B-splines are non-zero. This means that computational implementations
of B-splines and finite elements share similar aspects: matrix elements
are computed in batches on the individual knot intervals; only the
indexing of the basis functions is different. B-splines and their
applications to atomic and molecular physics have been reviewed by
Bachau in \citeref{Bachau2001}, to which I refer for further details.

The DVR,\citep{Harris1965,Dickinson1968,Lill1982,Light1985,Baye1986,Lill1986,Light2000}
in turn, is conventionally based on orthogonal polynomial bases and
their corresponding quadrature rules,\citep{Lill1982,Light1985} underlining
a strong connection to the finite element and B-spline approaches.
Generalizations of DVR to non-polynomial bases have been suggested
by Szalay\citep{Szalay1996,Szalay2003,Szalay2006,Szalay2013} and
Littlejohn, Cargo and coworkers\citep{Cargo2002,Cargo2002a,Littlejohn2002,Littlejohn2002a,Littlejohn2002b,Littlejohn2002c}.
The DVR community discerns between a variational basis representation
(VBR), in which all operator matrix elements \emph{i.e.} integrals
are computed exactly \emph{i.e.} analytically, a finite basis representation
(FBR), in which a quadrature rule is employed to approximate the integrals
like in FEM, and the DVR itself in which a unitary transformation
of the FBR basis set, which is often delocalized, is used to obtain
an effectively localized basis which affords a diagonal approximation
for matrix elements as
\begin{align}
O_{ij} & =\int_{a}^{b}\chi_{i}^{*}(x)O(x)\chi_{j}(x)dx\approx O(x_{i})\delta_{ij};\label{eq:O-diag}
\end{align}
this is the famous ``discrete $\delta$ property'' of DVR. Although
the DVR approach offers few benefits for one-dimensional problems
over the full, variational treatment, since cost of the additional
transformation has similar scaling to the savings, the approach becomes
extremely powerful in many dimensions due to the diagonality of potential
operators, \eqref{O-diag}.

The DVR is commonly used on top of a finite element treatment, yielding
the FEM-DVR method, as discussed e.g. by Rescigno and McCurdy.\citep{Rescigno2000}
Indeed, the DVR approaches discussed in the rest of the manuscript
all employ Lobatto elements, which use a basis of Lagrange interpolating
polynomials defined by a set of Gauss--Lobatto quadrature nodes,
as was discussed above in relation to FEM. It is easy to see that
if the same Gauss--Lobatto quadrature rule is used both to define
the finite element basis and for evaluating matrix elements, then
the property of the interpolating polynomials, \eqref{fem}, directly
leads to the diagonal approximation of matrix elements, \eqref{O-diag}.

However, this is an approximation: reliable computation of the matrix
elements $O_{ij}$ in \ref{eq:O-diag} may require using more than
$N$ quadrature points, which in turn may introduce off-diagonal elements
in the matrix representation of $\boldsymbol{O}$. Because of this
quadrature error, which is analogous to the aliasing error discussed
above in the case of finite-difference calculations, DVR calculations
can be non-variational; however, the error can be made smaller and
smaller by going to a larger DVR basis sets.

\section{Applications\label{sec:Applications}}

\subsection{Overview on general approaches for arbitrary molecules\label{subsec:Overview-on-general}}

With the ability to use varying levels of accuracy for different parts
of the system in the real-space basis set, even challenging parts
of the wave function, such as at the nucleus or far away from it,
can be straightforwardly converged to arbitrary accuracy. Real-space
methods thereby grant straightforward access even to properties that
are difficult to calculate accurately with \emph{e.g. }Gaussian basis
sets, such as nuclear magnetic properties, the electron density close
to, or at the nucleus, as well as the electron momentum density. A
further benefit of real-space methods is that the locality of the
real-space basis set allows for easy and efficient use of the algorithms
even on massively parallel architectures, whereas LCAO calculations
are not as nearly straightforward to parallellize to thousands of
processes.

Achieving the CBS limit is simple in real-space methods. Due to the
local support of the basis functions, the basis set never develops
linear dependencies, so the CBS limit can be achieved simply by increasing
the accuracy of the real-space basis set until the result does not
change anymore. The drawback to the arbitrary accuracy is that the
number of basis functions necessary for achieving even a \emph{qualitative}
level of accuracy in real-space approaches is significantly larger
than in the LCAO approach; the crossover typically happens only at
high precision.

The number of basis functions in real-space methods is especially
large in all-electron calculations, as disparate length scales are
introduced by the core orbitals. The description of the tightly bound
core orbitals necessitates an extremely robust basis near the nuclei,
whereas a much coarser basis is sufficient for the valence electrons.
One option to circumvent this problem is to forgo all-electron calculations
and to approximate the core electrons with pseudopotentials\citep{Hamann1979,Kleinman1982,Hamann1989,Pickett1989,Vanderbilt1990,Laasonen1993,Dolg2012}
or projector-augmented waves\citealp{Blochl1994} (PAWs), making the
problem accessible to a number of fully numerical approaches; see
\emph{e.g.} \citerefs{Arias1999, Beck2000, Torsti2006, Saad2010, Frediani2015}
for references. However, as the present review focuses on all-electron
calculations, the use of projector-augmented waves and pseudopotentials
will not be considered in the rest of the manuscript.

In order to make all-electron a.k.a. full-potential calculations feasible,
a suitable representation has to be picked for the description of
the core electrons. As was stated above, LCAO calculations have problems
especially with diffuse parts of the wave function, wheras core orbitals
don't pose significant complications. Vice versa, real-space methods
tend to struggle with core orbitals, but are well-adapted to the description
of diffuse character. This has lead to approaches that hybridize aspects
of LCAO and real-space calculations.\citep{Becke1990a,Pahl2002,Yamakawa2005,Rescigno2005,Yip2008,Talman2010,Losilla2012,Yip2014,Jerke2015,Solala2017,Parkkinen2017,Parkkinen2018,Kanungo2017,Marante2017,Solala2018}
By representing the core electrons using an atom-centric radial description,
a significantly smaller real-space basis set may suffice, making the
approaches scalable to large systems while still maintaining a very
high level of accuracy. In a somewhat similar spirit, the multi-domain
finite element muffin-tin and full-potential linearized augmented
plane-wave + local-orbital approaches have also been recently shown
to afford high-accuracy all-electron calculations for molecules.\citep{Braun2017,Gulans2018}

However, several methods that are able to solve the all-electron problem
without using an LCAO component exist as well. First, the FEM allows
for calculations at arbitrary precision even at complicated molecular
geometries by employing spatially non-uniform basis functions. A number
of FEM implementations for all-electron HF or DFT calculations on
molecules or solids with arbitrary geometries have been reported in
the literature.\citep{White1989,Pask1999,Pask2001,Pask2005,Tsuchida1995,Tsuchida1995a,Tsuchida1996,Tsuchida1998,Bylaska2009,Lehtovaara2009,Alizadegan2010,Suryanarayana2010,Lehtovaara2011,Motamarri2013,Schauer2013,Maday2014,Tsuchida2015,Schauer2015,Davydov2016}
Another solution is to use grids with hierarchical precision\citep{Wang2000,Cohen2013}
or adaptive multiresolution multiwavelet approaches\citep{Arias1999,Yanai2004,Yanai2004a,Yanai2005,Sekino2008,Sekino2012,Jensen2016,Jensen2017}
that are under active development and which have recently achieved
microhartree-accuracy in molecular calculations, even at the post-HF
level of theory.\citep{Bischoff2012,Bischoff2013,Kottmann2015,Yanai2015}
Regularization of the nuclear cusp\citep{Braun2012,Bischoff2014,Bischoff2014a}
and approaches employing adaptive coordinate systems\citep{Modine1997}
also have been suggested.

The considerably larger number of basis functions in real-space approaches
compared to LCAO calculations means not only that more computational
resources -- especially memory and disk space -- are needed, but
also that alternative approaches for \emph{e.g.} the solution of the
SCF equations have to be used. For instance, employing 100 degrees
of freedom such as grid points per Cartesian axis yields $(100)^{3}=10^{6}$
basis functions in total. The corresponding Fock matrix would formally
have a dimension $10^{6}\times10^{6}$, requiring a whopping 8 terabytes
of memory (assuming double precision), and making its exact diagonalization
untractable. Naturally, these challenges become even worse for molecules
for which orders of magnitude more points are necessary.

Furthermore, the calculation of properties may be complicated in real-space
methods. If the numerical basis set is independent of geometry, there
is no basis set superposition error as in LCAO calculations. Instead,
one has to deal with the infamous ``egg-box effect'',\citep{Artacho2008}
which is caused by the use of a finite spatial discretization that
breaks the translational and rotational symmetries of the system.
Then, the numerical accuracy depends on the relative position of the
nuclei to the discretization, which can manifest as \emph{e.g. }spurious
forces.\citep{Artacho2008} The egg-box effect can be made small by
going to larger and larger basis sets; other alternatives have been
discussed in \citeref{Ryu2016}, but the best approach may depend
on the form of the numerical basis set.

Typical real-space approaches employ local basis functions such as
the ones in \figref{Hat-functions.}, making most matrices extremely
sparse. Large-scale computational implementations are possible only
via the exploitation of this sparsity. For instance, the Coulomb interaction
as well as local and ``semilocal'' density functionals require only
local information, namely, the Coulomb potential $V_{\text{C}}(\boldsymbol{r})$
at every grid point, as well as the exchange-correlation potential
$V_{\text{XC}}(\boldsymbol{r})$ that is built from the density (and
possibly its gradient) at every grid point. In contrast, the exact
Hartree--Fock exchange interaction is non-local; however, it can
be truncated to finite range in large systems, especially if exact
exchange is only included within a short range, as in the popular
Heyd--Scuseria--Ernzerhof functional.\citep{Heyd2003,Heyd2006}
Alternatively, the evaluation of exact exchange can be sped up with
projection operators;\citep{Duchemin2010,Boffi2016} or the Krieger--Li--Iafrate
(KLI) approximation\citep{Krieger1990,Krieger1992} can be used to
build a fully local exchange potential, also allowing exact-exchange
calculations on large systems.\citep{Kim2015c,Lim2016a,Kim2017a,Kim2018}
Further necessary steps in the implementation of large-scale real-space
approaches involve replacing the full matrix diagonalization in the
conventional Roothaan orbital update of \eqref{Roothaan} with an
iterative approach such as the Davidson method\citealp{Davidson1975}
that is used to solve only for the new occupied subspace, or avoiding
diagonalization altogether by using \emph{e.g.} Green's function methods
to solve for the new occupied orbitals, as in the Helmholtz kernel
method.\citealp{Harrison2004,Frediani2013,Cools2014,Solala2017}

So far, the discussion has been completely general. Next, I shall
discuss calculations on atoms and diatomic molecules. Although it
is possible to obtain accurate results for atoms or diatomic molecules
by employing general three-dimensional real-space approaches as the
ones discussed above -- see \emph{e.g.} \citerefs{Ackermann1994}
and \citenum{Ackermann1995} for tetrahedral-element calculations
on \ce{H2+}, and \ce{H-} and \ce{He}, respectively -- they carry
a higher computational cost. A specialized implementation is able
to treat some dimensions analytically or semianalytically, whereas
these dimensions need to be fully described in the general three-dimensional
programs, necessitating orders of magnitude more degrees of freedom
to achieve the same level of accuracy as is possible in the specialized
algorithms.

For instance, while a $100^{3}$ grid can be assumed to yield a decent
HF or KS energy for any atom, an atomic solver will likely give a
much better energy with just $1000$ basis functions, as the angular
expansion for $s$ to $f$ electrons requires but 16 spherical harmonics,
while 60 radial functions afford sub-microhartree accuracy even for
xenon.\citep{Lehtola2019a} Furthermore, the smaller basis set afforded
by the use of the symmetry inherent in the problem allows (di)atomic
calculations to use a diversity of powerful approaches -- such as
the use of \eqref{Roothaan} with full diagonalization for the orbital
update -- which are not tractable in the general three-dimensional
approach.\cite{Lehtola2019a, Lehtola2019b}

\subsection{Atoms\label{subsec:Atoms}}

At the non-relativistic level of theory, the electronic Hamiltonian
for an atom with nuclear charge $Z$
\begin{equation}
\hat{H}_{el}=-\frac{1}{2}\sum_{i}\nabla_{i}^{2}-\sum_{i}\frac{Z}{r_{i}}+\sum_{i>j}\frac{1}{r_{ij}}\label{eq:atH}
\end{equation}
commutes with the total angular momentum operator $\hat{\boldsymbol{L}}^{2}$,
its $z$ component $\hat{L}_{z}$, as well as the total spin operator
$\hat{\boldsymbol{S}}$. As a result, electronic states are identified
with the term symbol $^{2S+1}L$, where the total electronic angular
momentum $L$ is one of $S,P,D,F,G\dots$. In cases where the term
symbol is not $^{n}S$, the wave function is in principle strongly
correlated due to the degeneracy of partially filled shells. Unlike
the molecular case, in which the treatment of strong correlation effects
is still under active research, for atoms strong correlation effects
are much more docile due to the limited number of accessible orbitals.
For example, in the ground state configuration of boron the $2p$
shell only has a single electron, which should be distributed equally
between the $p_{x}$, $p_{y}$ and $p_{z}$ orbitals, yielding a multiconfigurational
(MC) wave function. A good symmetry-compatible ansatz for the wave
function can be constructed with Russell--Saunders LS-coupling,\citep{Sturesson1993}
yielding \emph{e.g.} the MCHF method. Such calculations can also be
formulated as an average over the energies of the individual configurations.\citep{Roothaan1960,Huzinaga1960,Huzinaga1961,Huzinaga1969a,Davidson1973}

Fully numerical calculations are possible for atoms, because the two-electron
interaction has a compact representation, as shown by the Legendre
expansion
\begin{equation}
\frac{1}{r_{12}}=\frac{4\pi}{r_{>}}\sum_{L=0}^{\infty}\frac{1}{2L+1}\left(\frac{r_{<}}{r_{>}}\right)^{L}\sum_{M=-L}^{L}Y_{L}^{M}(\Omega_{1})\left(Y_{L}^{M}(\Omega_{2})\right)^{*}.\label{eq:legendre-exp}
\end{equation}
The numerical representation of the atomic orbitals was already given
in \eqref{NAO}, which yields the boundary value $u_{nl}(r)=0$ for
all values of $l$. It is seen that the two-electron integrals 
\begin{align}
(ij|kl)= & \iint\frac{\chi_{i}(\boldsymbol{r})\chi_{j}^{*}(\boldsymbol{r})\chi_{k}(\boldsymbol{r}')\chi_{l}^{*}(\boldsymbol{r}')}{\left|\boldsymbol{r}-\boldsymbol{r}'\right|}{\rm d}^{3}r{\rm d}^{3}r'\label{eq:tei}
\end{align}
reduce to an angular part times a radial part in such a basis set.
The angular part can be evaluated analytically using the algebra of
spherical harmonics, whereas the radial parts involve but one- and
two-dimensional integrals, which can be readily evaluated by quadrature.

Indeed, real-space calculations on atoms in the (MC)HF approach have
a long history, starting out with finite-difference approaches.\citep{Froese1963,FroeseFischer1970,FroeseFischer1972a,Chernysheva1976,FroeseFischer1978,Biegler-Konig1986,FroeseFischer1987,FroeseFischer1991b,Andrae1997,FroeseFischer2000,FroeseFischer2007}
By employing large grids, the atomic energies and wave functions can
be straightforwardly converged to the complete basis set limit. The
MCHF energies for atoms have been known to high precision for a long
time;\citep{FroeseFischer1968,FroeseFischer1972b,FroeseFischer1973b,Koga1995,Koga1995a,Koga1996,Saito2009}
reproduction of the exact ground state energy, however, is a hard
problem even for the Li atom.\citep{King1997} Ivanov and Schmelcher
have employed finite differences to study atoms in weak to strong
magnetic fields.\citep{Ivanov1994,Ivanov1998,Ivanov1999a,Ivanov2000,Ivanov2001,Ivanov2001a,Ivanov2001b}

More recently, finite difference calculations have been complemented
with FEM\citep{Flores1989,Flores1989a} as well as B-spline approaches.\citep{Gilbert1974,Gilbert1974a,Gilbert1975,Gilbert1975a,AltenbergerSiczek1976,Gazquez1977,Silverstone1978,Carroll1979,FroeseFischer1990,FroeseFischer1992,Hansen1993,FroeseFischer1995a,Qiu1999,Saito2003,FroeseFischer2008,FroeseFischer2011}
In addition to real-space methods, also momentum space approaches
have been suggested,\citep{Delhalle1987,Defranceschi1989,Defranceschi1990,Defranceschi1992,DeWindt1992,DeWindt1993,Windt1993,DeWindt1994,Fischer1994}
but they have not become widely used. As was discussed in \secref{Real-space-methods},
the finite element and B-spline methods are similar to each other:
either one can be used for a systematic and smooth approach to the
complete basis set limit; also the implementations of the two approaches
are similar. In addition to enabling variational calculations --
approaching the converged value strictly from above -- and making
sure orbital orthonormality is always honored, the use of an explicit
radial basis set instead of the implicit basis set of the finite-difference
method also enables the straightforward use of post-HF methods.\citep{FroeseFischer2016}
For instance, Flores \emph{et al.} have calculated correlation energies
with Møller--Plesset perturbation theory\citep{Moller1934} truncated
at the second order,\citep{Flores1992,Flores1992a,Flores1993,Flores1993a,Flores1993b,Flores1994,Flores1997,Flores1999,Flores2003,Flores2004,Jankowski2004a,Flores2006,Flores2008}
while Sundholm and coworkers have relied on highly accurate multiconfigurational
SCF methods\citep{Sundholm1989} to study the ground state of the
beryllium atom,\citep{Martensson-Pendrill1991} electron affinities,\citep{Sundholm1990,Sundholm1994,Olsen1994a,Sundholm1995,Sundholm1995b}
excitation energies and ionization potentials,\citep{Sundholm1991c,Sundholm1992,Sundholm1993d}
hyperfine structure,\citep{Sundholm1990a,Sundholm1991,Sundholm1991a}
nuclear quadrupole moments,\citep{Sundholm1990b,Sundholm1991a,Sundholm1991b,Sundholm1992a,Sundholm1992b,Sundholm1992c,Sundholm1993,Sundholm1993a,Sundholm1994a,Olsen1994,Tokman1997,Tokman1998,Sundholm1999,Kello1999,Bieron2001,Sundholm2018}
the extended Koopmans' theorem,\citep{Sundholm1993b,Sundholm1993c,Olsen1998}
and the Hiller--Sucher--Feinberg identity.\citep{Sundholm1995a}
Braun\citep{Braun2002} and by Engel and coworkers\citep{Engel2008,Engel2009,Schimeczek2014}
have in turn reported finite-element calculations of atoms in strong
magnetic fields. Approaches based on FEM-DVR have also been developed:
Hochstuhl and Bonitz have implemented a restricted active space procedure
for modeling photoionization of many-electron atoms.\citep{Hochstuhl2012}
For more details and references on multiconfigurational calculations
on atoms, I refer to the recent review by Froese Fischer and coworkers
in \citeref{FroeseFischer2016}.

As is obvious from the above references, multiple programs are available
for atomic calculations at the HF and post-HF levels of theory. However,
the situation is not as good for DFT. Note that unlike wave function
based methods such as MCHF, in theory a single-determinant approach
should suffice for DFT also in the case of significant strong correlation
effects.\citealp{Kohn1965} Unfortunately, this is true only for the
(unknown) exact exchange-correlation functional, while approximate
functionals typically yield unreliable estimates for strongly correlated
molecular systems;\citep{Becke2014,Jones2015,Mardirossian2017a} this
is an active research topic within the DFT community.\citealp{Malet2012,Becke2013,Becke2013a,Johnson2013a,Gagliardi2016,Li2016,Johnson2017,Su2018a}

Before gaining acceptance among chemists, DFT was popular for a long
time within the solid state physics community. As was already stated
above, solid state calculations typically employ pseudopotentials\citep{Hamann1979,Kleinman1982,Hamann1989,Pickett1989,Vanderbilt1990,Laasonen1993,Dolg2012}
or projector-augmented waves\citep{Blochl1994} (PAWs) to avoid explicit
treatment of the highly localized and chemically inactive core electrons.\citep{Schwerdtfeger2011}
Even though DFT calculations on atoms are the starting point for the
construction of pseudopotentials and PAW setups -- meaning that a
number of atomic DFT programs such as fhi98PP,\citep{Fuchs1999} atompaw,\citep{Holzwarth2001}
APE,\citep{Oliveira2008} as well as one in GPAW\citep{Larsen2009}
are commonly available -- the scope of these programs appears to
be limited, restricting their applicability for general calculations.

Especially, atomic DFT programs typically employ spherically averaged
occupations
\begin{equation}
n(r)=\sum_{nl}f_{nl}\sum_{m}|\psi_{nlm}(\boldsymbol{r})|^{2}=\sum_{nl}\frac{\left(2l+1\right)f_{nl}}{4\pi}\left(\frac{u_{nl}(r)}{r}\right)^{2},\label{eq:sphave}
\end{equation}
which allows for huge performance benefits, since now calculations
employing the local spin density approximation\citep{Kohn1965} (LDA)
or the generalized gradient approximation\citep{Langreth1980} (GGA)
can be formulated as a set of coupled one-dimensional radial Schrödinger
equations that can be solved in a matter of seconds even in large
grids. Using such an approach, for instance Andrae, Brodbeck, and
Hinze have investigated closed-shell $^{1}S$ states for $2\ensuremath{\leq Z\leq102}$
to study differences between HF and DFT approaches.\citep{Andrae2001a}

Unfortunately, the restriction to spherical symmetry prevents the
application of such programs to e.g. the prediction of electric properties,
as electric fields break spherical symmetry. Furthermore, due to the
computational and theoretical challenges of using exact exchange in
solid state calculations, support for hybrid exchange-correlation
functionals is limited in the existing atomic DFT programs. To my
best knowledge, only one atomic program that supports both HF and
DFT calculations has been described in the literature,\citep{Ozaki2011}
but it only supports calculations at LDA level, lacks hybrid functionals,
and employs but cubic Hermite basis functions. Another program employing
higher-order basis functions has been reported, but it is again restricted
to LDA calculations and also lacks support for HF.\citep{Romanowski2008,Romanowski2009,Romanowski2009a,Romanowski2009b}
However, as discussed in \subsecref{Self-consistent-field-approaches},
I have recently solved this issue in \citeref{Lehtola2019a}.

\subsection{Diatomic molecules\label{subsec:Diatomic-molecules}}

At the non-relativistic level of theory, the Hamiltonian for the electrons
in a diatomic molecule composed of nuclei with charges $Z_{A}$ and
$Z_{B}$ separated by $R_{AB}$ is given by
\begin{equation}
\hat{H}_{el}=-\frac{1}{2}\sum_{i}\nabla_{i}^{2}-\sum_{i}\frac{Z_{A}}{r_{iA}}-\sum_{i}\frac{Z_{B}}{r_{iB}}+\sum_{i>j}\frac{1}{r_{ij}}+\frac{Z_{A}Z_{B}}{R_{AB}}\label{eq:diatH}
\end{equation}
and the electronic states are identified in analogy to atomic states
with the term symbol $^{2S+1}\Lambda$, where the total electronic
angular momentum quantum number measured along the internuclear axis
$\Lambda$ is one of $\Sigma,\Pi,\Delta,\Phi,\Gamma,\dots$, with
strong correlation again arising via degenerate levels especially
when the term symbol is not $\Sigma$. Although diatomic molecules
are not as straightforward for numerical treatment as atoms due to
the existence of two nuclei with two nuclear cusps, it was noticed
early on by McCullough that a fully numerical treatment is tractable
via the choice of a suitable coordinate system.\citep{McCullough1974,McCullough1975,McCullough1986}

\begin{figure}
\begin{centering}
\includegraphics[width=0.5\textwidth]{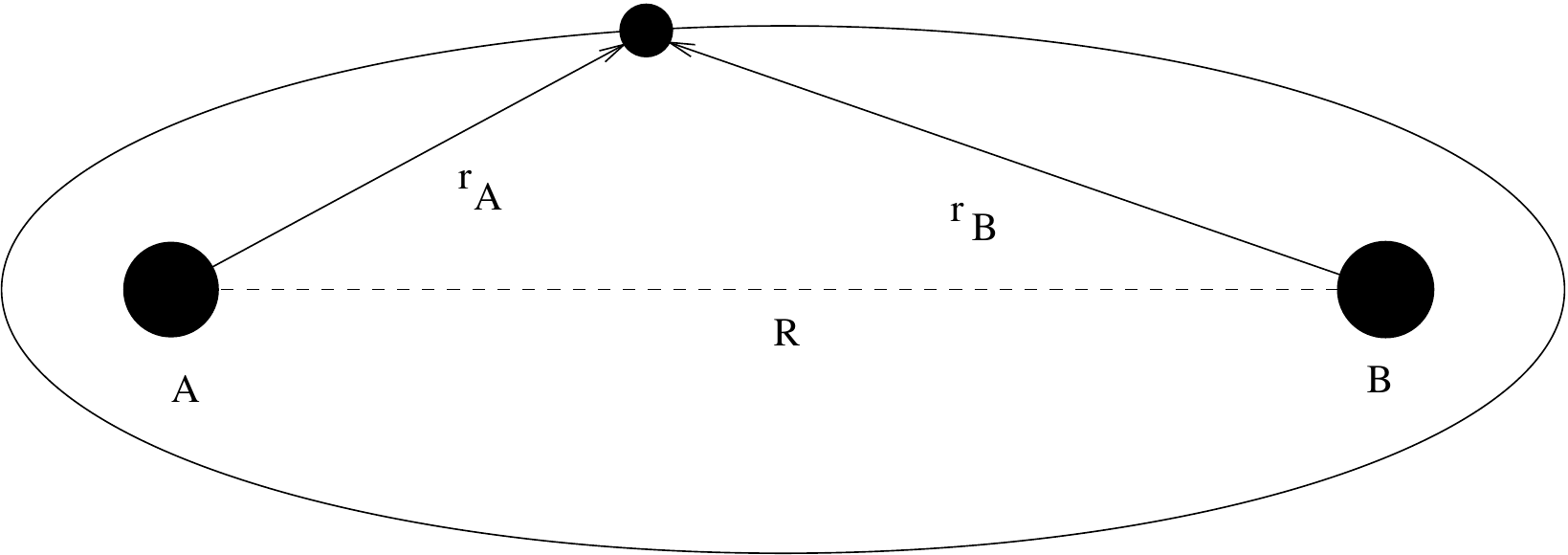}
\par\end{centering}
\caption{Illustration of the diatomic coordinate system defined by nuclei A
and B, depicted by the large balls. The position of an electron (small
ball) is parametrized in terms of distances $r_{A}$ and $r_{B}$
from the nuclei A and B. The ellipse with foci at A and B describes
an isosurface of $\xi$, which is drawn by $\eta\in[-1,1]$.\label{fig:coordsys}}
\end{figure}

\begin{figure}
\begin{centering}
\includegraphics[width=0.5\textwidth]{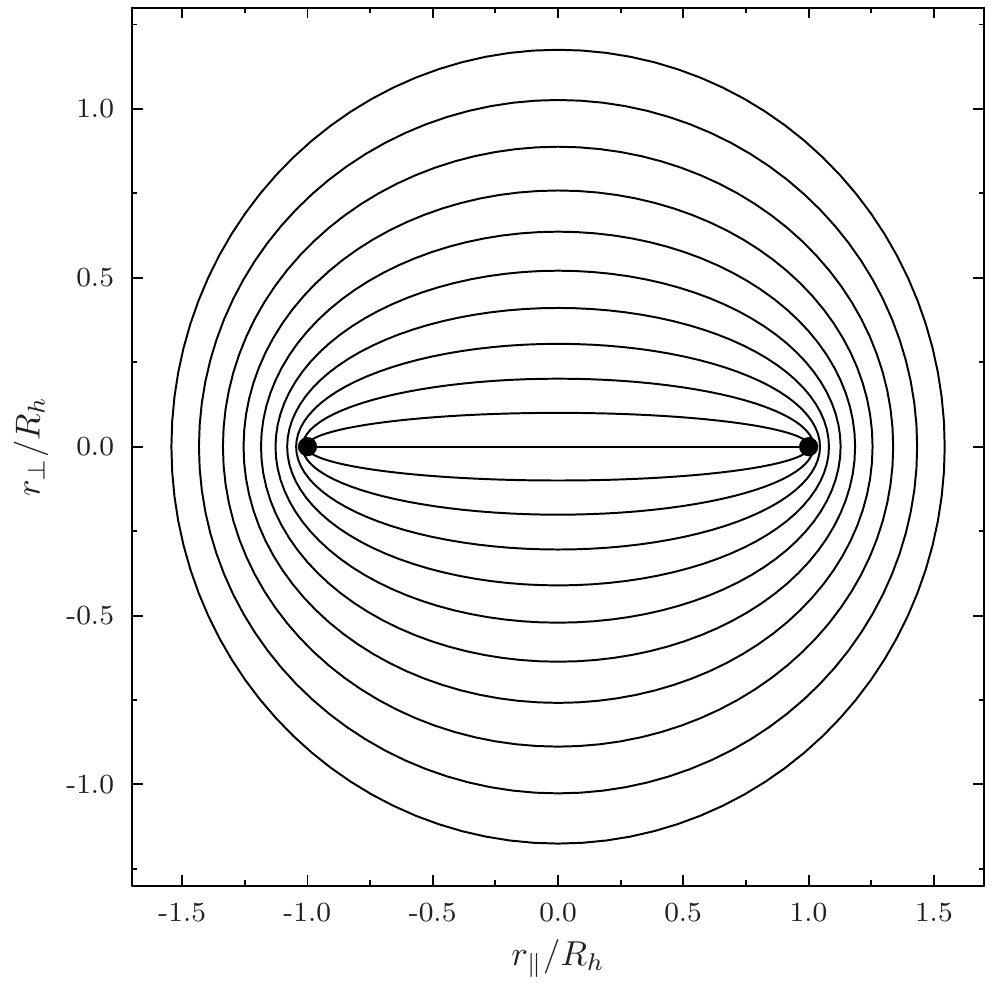}
\par\end{centering}
\caption{Illustration of isosurfaces of $\xi$ in terms of distances parallel
to the bond (along the molecular $z$ axis) and perpendicular to the
bond (in the molecular $(x,y)$ plane). The isosurface for $\xi=1$
is the straight line connecting the nuclei, with larger values of
$\xi$ corresponding to larger and larger ellipses. For $\xi\to\infty$
the isosurfaces become circles. \label{fig:isosurface-mu}}
\end{figure}

\begin{figure}
\begin{centering}
\includegraphics[width=0.5\textwidth]{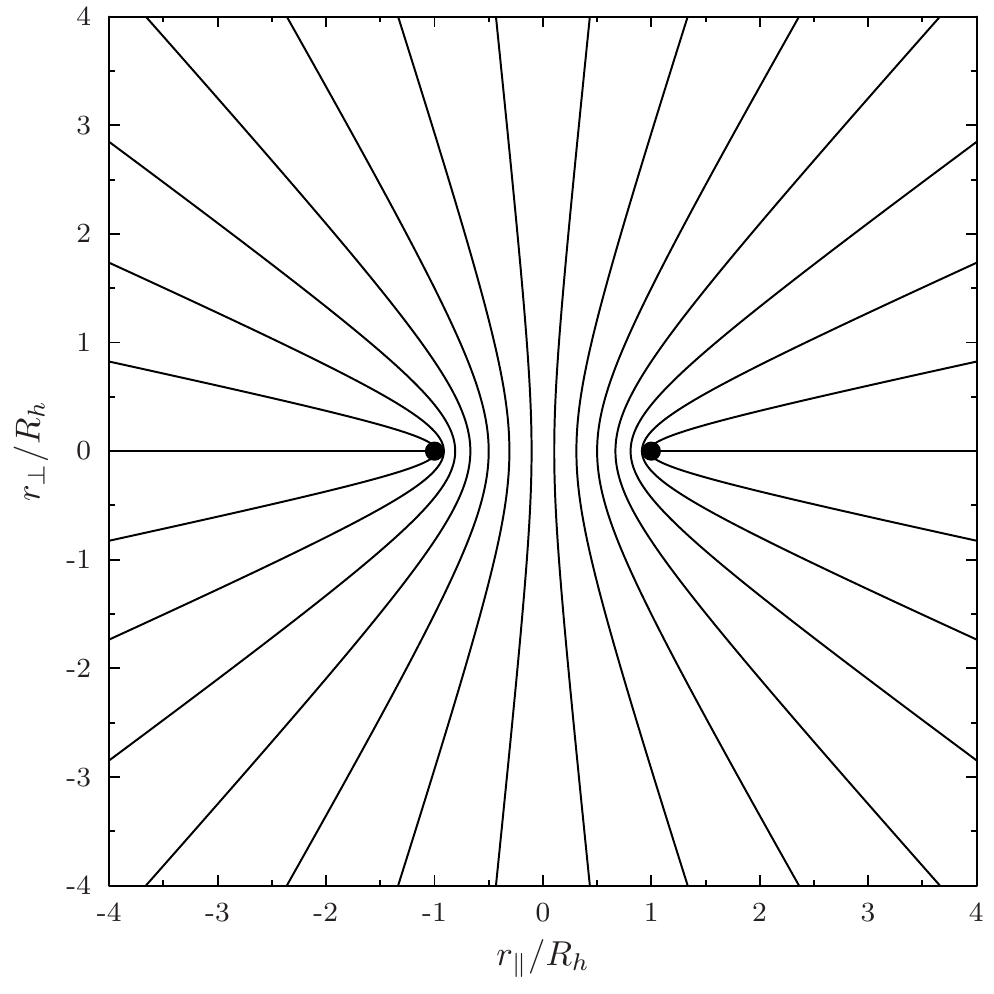}
\par\end{centering}
\caption{Illustration of isosurfaces of $\eta$ in terms of distances parallel
to the bond (along the molecular $z$ axis) and perpendicular to the
bond (in the molecular $(x,y)$ plane). The isosurfaces for $\eta=-1$
and $\eta=+1$ are the straight lines emanating from the nuclei towards
towards $-\infty$ and $+\infty$, respectively, with intermediate
values of $\eta$ corresponding to the curved isosurfaces. For $\eta=0$
the isosurface is the plane $(x,y,z=0)$. \label{fig:isosurface-nu}}
\end{figure}

Analogously to atoms, where the orbitals can be written in the form
of \eqref{atorb}, placing the nuclei along the $z$ axis the orbitals
for a diatomic molecule become expressible as
\begin{equation}
\psi_{nm}(\boldsymbol{r})=\chi_{nm}(\xi,\eta)e^{im\phi};\label{eq:diatom}
\end{equation}
the same blocking by $m$ holds for any linear molecule. The $m$
quantum number describes the orbital character (see Appendix A for
a brief discussion), and the prolate spheroidal coordinates are given
by
\begin{align}
\xi= & \frac{r_{A}+r_{B}}{R},\;1\leq\xi<\infty,\label{eq:xi}\\
\eta= & \frac{r_{A}-r_{B}}{R},\;-1\leq\eta\leq1.\label{eq:eta}
\end{align}
with $\phi$ measuring the angle around the bond axis. The coordinate
system given by \eqref{xi, eta} is illustrated in \figref{coordsys, isosurface-mu, isosurface-nu}.
Even though the $\xi$ and $\eta$ coordinates in the $(\xi,\eta,\phi)$
prolate spheroidal coordinate system don't have as clear roles as
$(r,\theta,\phi)$ in spherical polar coordinates, as can be seen
from \figref{isosurface-mu, isosurface-nu}, the prolate spheroidal
coordinate $\xi$ can still be roughly identified as a ``radial''
coordinate, whereas $\eta$ can be identified as an ``angular''
coordinate that describes variation along the bond direction.

As in the spherical polar coordinate system for atoms, the nuclear
attraction integrals have no singularities in the $(\xi,\eta,\phi)$
coordinate system in the case of diatomic molecules, allowing for
smooth and quick convergence to the CBS limit. As in the case of atoms,
where the volume element $r^{2}{\rm d}\Omega$ cancels the $r^{-1}$
divergence of the nuclear Coulomb attraction, the volume element of
the prolate spheroidal coordinate system\citep{Rudenberg1951}
\begin{equation}
{\rm d}V=\frac{1}{8}R^{3}\left(\xi^{2}-\eta^{2}\right){\rm d}\xi{\rm d}\eta{\rm d}\phi\label{eq:dV-1}
\end{equation}
can be rewritten in the form
\begin{equation}
{\rm d}V=\frac{1}{2}R\ r_{A}(\xi,\eta)r_{B}(\xi,\eta)\ {\rm d}\xi{\rm d}\eta{\rm d}\phi\label{eq:dV-2}
\end{equation}
where the product $r_{A}r_{B}$ kills off the divergences of $r_{A}^{-1}$
and $r_{B}^{-1}$ in attraction integrals. Equally importantly, the
two-electron integrals have a compact representation in the prolate
spheroidal coordinate system due to the Neumann expansion\citep{Rudenberg1951}
\begin{align}
\frac{1}{r_{12}}= & \frac{8\pi}{R}\sum_{L=0}^{\infty}\sum_{M=-L}^{L}(-1)^{M}\frac{\left(L-|M|\right)!}{\left(L+|M|\right)!}P_{L}^{|M|}(\xi_{<})Q_{L}^{|M|}(\xi_{>})Y_{L}^{M}(\Omega_{1})\left(Y_{L}^{M}(\Omega_{2})\right)^{*}.\label{eq:neumann}
\end{align}
As is obvious from \eqrangeref{xi}{neumann}, the coordinate system
depends on the internuclear distance. This means that there are no
egg-box effects in the diatomic calculations; however, the accuracy
of the $(\xi,\eta)$ description now depends on the geometry of the
system, as discussed in \citeref{Lehtola2019b}.

In McCullough's pioneering approach, the orbitals are expanded as\citep{McCullough1974}
\begin{equation}
\chi_{nm}(\xi,\eta)=\sum_{l=|m|}^{\infty}f_{nml}(\xi)P_{l}^{m}(\eta),\label{eq:McCullough}
\end{equation}
where $f_{nml}(\xi)$ are unknown functions determined on a grid using
finite differences, and $P_{l}^{m}$ are normalized associated Legendre
functions. Alternatively, rearranging \eqref{diatom,McCullough} the
expansion can be written as\citep{McCullough1975} 
\begin{equation}
\psi_{nm}(\boldsymbol{r})=\sum_{l=|m|}^{\infty}X_{nml}(\xi)Y_{l}^{m}(\eta,\phi),\label{eq:McCullough-2}
\end{equation}
where $X_{nml}(\xi)=f_{nml}(\xi)/\sqrt{4\pi}$. McCullough called
the approach ``the partial-wave self-consistent field method'' (PW-SCF),
as in practice only a finite number of partial waves $l$ can be included
in the expansion. Pioneering work by McCullough and coworkers applied
the PW-SCF method to the study of parallel polarizabilities\citep{McCullough1975a,Christiansen1977,Christiansen1979}
and the accuracy of LCAO calculations.\citep{Christiansen1977a} The
work on the approach continued with multiconfigurational self-consistent
field (MCSCF) calculations\citep{Adamowicz1981,McCullough1982} and
excited states,\citep{McCullough1981,Adamowicz1984a} quadrupole moments,\citep{McCullough1981a}
post-HF methods,\citep{McCullough1984,Adamowicz1985} the extended
Koopmans' theorem,\citep{Adamowicz1986} hyperfine splitting constants,\citep{Richman1987}
the chromium\citep{Richman1987a} and copper dimers,\citep{Partridge1988}
as well as magnetic hyperfine parameters.\citep{Beck1990} Adamowicz
and coworkers used the program for post-HF methods\citep{Adamowicz1984c,Adamowicz1988b,Adamowicz1989}
and electric polarizabilities,\citep{Adamowicz1988a} and Chipman
used it for calculating spin densities.\citep{Chipman1989}

Contemporaneously, motivated by McCullough's initial work, Laaksonen,
Sundholm and Pyykkö pursued an approach in which $\chi_{nm}(\xi,\eta)$
in \eqref{diatom} is determined directly with a finite difference
procedure.\citep{Laaksonen1983,Laaksonen1983a} At the same time,
Becke developed a program for LDA calculations on diatomic molecules,\citep{Becke1982,Becke1983,Becke1985,Becke1986,Becke1986b}
proposing the use of a further coordinate transform
\begin{align}
\xi=\cosh\mu & ,\;0\leq\mu<\infty,\label{eq:mu}\\
\eta=\cos\nu & ,\;0\leq\nu\leq\pi,\label{eq:nu}
\end{align}
which eliminates nuclear cusps in the wave function. This can be easily
seen by solving \eqref{xi,eta} for the distances from the nuclei
and by substituting Taylor expansions of \eqref{mu,nu} around $(\mu,\nu)=(0,0)$,
and around $(\mu,\nu)=(0,\pi)$ with $\Delta\nu=\nu-\pi$:
\begin{align}
r_{A}\approx & R\left[1+\frac{1}{4}\mu^{2}-\frac{1}{4}\nu^{2}+{\cal O}(\mu^{4})+{\cal O}(\nu^{4})\right],\label{eq:ra-exp}\\
r_{B}\approx & R\left[1+\frac{1}{4}\mu^{2}-\frac{1}{4}\left(\Delta\nu\right)^{2}+{\cal O}(\mu^{4})+{\cal O}\left(\left(\Delta\nu\right)^{4}\right)\right],\label{eq:rb-exp}
\end{align}
Thus, while an exponential function $\exp\left(-\zeta r_{A}\right)$
centered on atom A has a cusp in both $\xi$ and $\eta$ in the $(\xi,\eta)$
coordinate system, there is no cusp in the $(\mu,\nu)$ coordinate
system. Thanks to \eqref{ra-exp}, the exponential turns into a Gaussian
function, which is much easier to represent numerically in both $\mu$
and $\nu$.

The coordinate transform of \eqref{mu,nu} was instantly adopted by
Laaksonen, Sundholm, and Pyykkö for HF,\citep{Laaksonen1983b,Laaksonen1988,Muller-Plathe1989,Muller-Plathe1989a,Laaksonen1986,Pyykko1987,Pyykko1987a,Sundholm1985}
MC-SCF,\citep{Laaksonen1984} and density functional calculations.\citep{Laaksonen1985,Sundholm1985b}
Laaksonen, Sundholm, Pyykkö, and others also presented applications
of the method to the calculation of electric field gradients,\citep{Sundholm1985a,Baerends1985}
nuclear quadrupole moments,\citep{Sundholm1984,Sundholm1986,Diercksen1988}
as well as repulsive interatomic potentials.\citep{Nordlund1997}
Note, however, that Becke's own work employed a further transform
from the $(\mu,\nu)$ coordinates, see \citeref{Becke1982}. 

The development of the program originally written by Laaksonen, Sundholm
and Pyykkö was taken over by Kobus,\citep{Kobus1996,Kobus2012,Kobus2013}
who proposed an alternative relaxation approach for solving the self-consistent
field equations,\citep{Kobus1994a} and extensively studied deficiencies
of LCAO basis sets,\citep{Kobus1994,Kobus1994b,Kobus1995,Moncrieff1995b,Kobus1997,Moncrieff1998a,Kobus1999,Kobus2000,Kobus2000a,Kobus2001,Kobus2001a,Kobus2001b,Kobus2002,Kobus2004,Kobus2007a,Glushkov2008,Styszynski2003,Matito2006a}
hydrogen halides,\citep{Styszynski2000,Styszynski2003} multipole
moments and parallel polarizabilities,\citep{Kobus2007,Kobus2015}
and molecular orbitals in a Xe-C ion-target system.\citep{Wilhelm2017}

The \textsc{x2dhf} program is open source and is still maintained,
and publicly available on the internet.\citep{x2dhf} \textsc{x2dhf}
has been used \emph{e.g.} by Jensen in the development of polarization
consistent basis sets,\citep{Jensen2001,Jensen2002,Jensen2002a,Jensen2002b,Jensen2003,Jensen2004}
which suggested that some numerical HF energies in the literature
were inaccurate due to an insufficient value for the practical infinity
$r_{\infty}$.\citep{Jensen2005} Namely, as the Coulomb and exchange
potentials are determined in \textsc{x2dhf}\citep{Laaksonen1986,Kobus1996,Kobus2013}
by relaxation approaches that start from asymptotic multipole expansions,\citep{Laaksonen1983b}
the practical infinity $r_{\infty}$ may need to be several hundred
atomic units away to reach fully converged results, even if the electron
density itself typically vanishes in a fraction of this distance.

The \textsc{x2dhf} program was also used by Grabo and Gross for optimized
effective potential calculations within the KLI approximation,\citep{Grabo1997,Grabo1997a}
by Karasiev and coworkers for density functional development,\citep{Liu1996,Liu1998,Ludena1999,Karasiev1999,Karasiev2000,Karasiev2002,Karasiev2002a,Karasiev2003,Karasiev2006}
by Halkier and Coriani for computing molecular electric quadrupole
moments,\citep{Halkier2001} Roy and Thakkar who studied MacLaurin
expansions of the electron momentum density for 78 diatomic molecules,\citep{Roy2002}
Weigend, Furche, and Ahlrichs for studying total energy and atomization
energy basis set errors in quadruple-$\zeta$ basis sets,\citep{Weigend2003}
Shahbazian and Zahedi who studied basis set convergence patterns,\citep{Shahbazian2005}
Pruneda, Artacho, and Kuzmin for the calculation of range parameters,\citep{Pruneda2004,Kuzmin2006,Kuzmin2007}
Madsen and Madsen for modeling high-harmonic generation,\citep{Madsen2006}
Williams \emph{et al.} who reported numerical HF energies for transition
metal diatomics,\citep{Williams2008} Madsen and coworkers who calculated
structure factors for tunneling,\citep{Madsen2012,Madsen2013,Saito2015,Madsen2017}
Kornev and Zon who studied anti-Stokes-enhanced tunneling ionization
of polar molecules,\citep{Kornev2014} and Endo and coworkers who
studied laser tunneling ionization of NO.\citep{Endo2016}

Based on the results of Laaksonen \emph{et al}., Kolb and coworkers
developed a FEM program for HF and density functional calculations
employing triangular basis functions, again in the $(\mu,\nu)$ coordinate
system.\citep{Schulze1985,Heinemann1987,Heinemann1988,Heinemann1988a,Heinemann1989,Heinemann1990,Heinemann1990a,Heinemann1993,Kopylow1998,Kopylow1998a}
A finite element variant employing rectangular elements was also developed
by Sundholm and coworkers for MCSCF calculations,\citep{Sundholm1988,Sundholm1989}
whereas a multigrid conjugate residual HF solver for diatomics has
been described by Davstad.\citep{Davstad1992} Makmal, Kümmel, and
Kronik studied fully numerical all-electron solutions to the optimized
effective potential equation with and without the KLI approximation,
also employing a finite difference approach.\citep{Makmal2009,Makmal2009-err}
Morrison \emph{et al.} have also discussed various numerical approaches
for diatomic molecules.\citep{Morrison2000,Morrison2016,Morrison2018}
Artemyev and coworkers have reported an implementation of the partial-wave
approach based on B-splines.\citep{Artemyev2004}

Apparently unaware of the unanimous agreement between Becke, Laaksonen,
Sundholm, Pyykkö, Kobus, Kolb, and others in the quantum chemistry
community on the supremacy of the $(\mu,\nu)$ coordinate system,
several works utilizing the original $(\xi,\eta)$ coordinates defined
by \eqref{xi,eta} have been published in the molecular physics literature.
For instance, a B-spline configuration interaction program has been
reported by Vanne and Saenz,\citep{Vanne2004} whereas FEM-DVR expansions
have been employed by Tao and coworkers\citep{Tao2009,Tao2010} as
well as Guan and coworkers\citep{Guan2010,Guan2011} for studying
dynamics of the \ce{H2} molecule, as well as by Tolstikhin, Morishita,
and Madsen for studying \ce{HeH^{2+}}.\citep{Tolstikhin2011} Haxton
and coworkers have reported multiconfiguration time-dependent HF calculations
of many-electron diatomic molecules with FEM-DVR basis sets,\citep{Haxton2011}
whereas Larsson and coworkers\citep{Larsson2016} and Yue and coworkers\citep{Yue2017}
have studied correlation effects in ionization of diatomic molecules
employing restricted-active space calculations also with FEM-DVR.
A FEM-DVR -based program supporting HF calculations for diatomics
has been published by Zhang and coworkers.\citep{Zhang2015} All of
these programs employ a product grid in $\xi$ and $\eta$. Momentum
space approaches have also been developed for diatomic molecules\citep{Berthier1985,Fischer1992a,Fischer1993a,Berthier1997}
as well as larger linear molecules,\citep{Defranceschi1984,Berthier1985,Defranceschi1986}
but they have not become widely used.

\section{Summary and outlook\label{sec:Summary-and-outlook}}

I have reviewed approaches for non-relativistic all-electron electronic
structure calculations based on the linear combination of atomic orbitals
(LCAO) employing a variety of radial functions, and discussed their
benefits and drawbacks. The shortcomings of LCAO calculations were
used to motivate a need for fully numerical approaches, which were
also briefly discussed along with their good and bad sides. I pointed
out that fully numerical approaches can be made more affordable for
atoms and diatomic molecules by exploiting the symmetry inherent in
these systems, and reviewed the extensive literature on fully numerical
quantum chemical calculations on atoms and diatomic molecules.

\subsection{Self-consistent field approaches\label{subsec:Self-consistent-field-approaches}}

The development of novel density functionals might especially benefit
from access to affordable, fully numerical electronic structure approaches
for atoms and diatomic molecules. As evidenced by the existence of
several numerically ill-behaved functionals,\citealp{Grafenstein2007,Johnson2009a,Wheeler2010,Mardirossian2013,Bircher2018}
it has not been common procedure to assess the behavior of new functionals
with fully numerical calculations. The flexibility of fully numerical
basis sets guarantees that any numerical instabilities in a functional
are instantly seen in the results of a calculation, in contrast to
calculations in small LCAO basis sets.

I have recently published a program for fully numerical electronic
structure calculations on atoms\citep{Lehtola2019a} and diatomic
molecules\citep{Lehtola2019b} for HF and DFT called \textsc{HelFEM}.\citep{HelFEM}
In the atomic program,\citep{Lehtola2019a} finite elements are used
for the radial $u_{nl}(r)$ expansion, whereas the diatomic program\citep{Lehtola2019b}
employs the partial wave approach of McCullough\citep{McCullough1974,McCullough1975}
to represent the angular $\nu$ and $\phi$ directions using spherical
harmonics, while one-dimensional finite elements are again used in
the radial $\mu$ direction as in the atomic case.\citep{Lehtola2019a}

The freely available, open-source \textsc{HelFEM} program introduces
support for hybrid density functionals, as well as meta-GGA functionals,
which have not been publicly available in other fully numerical program
packages for calculations on atoms or diatomic molecules; hundreds
of functionals are supported via an interface to the \textsc{Libxc}
library.\citealp{Lehtola2018} In addition to calculations in a finite
electric field,\citep{Lehtola2019a,Lehtola2019b} \textsc{HelFEM}
has been recently extended to calculations in finite magnetic fields
that are especially challenging to model with popular LCAO approaches.\citep{Lehtola2019d}
Due to the close coupling of \textsc{HelFEM} to \textsc{Libxc,} my
hope is that the program will be adopted for the development of new
density functionals, as both atoms and diatomic molecules can be calculated
easily and efficiently directly at the basis set limit.

Unlike previous programs relying on relaxation approaches to solve
the Poisson and HF/KS equations, the approach used in \textsc{HelFEM}
is fully variational, guaranteeing a monotonic convergence from above
to the ground state energy. Extremely fast convergence to the radial
grid limit is observed both for the atomic and diatomic calculations,
with less than 100 radial functions typically guaranteeing full convergence
even in the case of heavy atoms.

The angular expansions are typically small in the case of atoms, even
in the presence of finite electric fields that break spherical symmetry.\citep{Lehtola2019a}
However, in the case of diatomics, the angular expansions may converge
less slowly. Still, I have shown in \citeref{Lehtola2019b} that a
non-uniform expansion length in $m$ affords considerably cheaper
calculations with no degradation in accuracy. Although \citeref{Lehtola2019b}
shows that the partial wave expansion yields extremely compact expansions
for a large number of molecules as compared to finite-difference calculations
with \textsc{x2dhf} of the same quality, going beyond fourth period
elements may require further optimizations and parallellization in
the \textsc{HelFEM} program.

The approach used in \textsc{HelFEM} is fast for light molecules that
do not require going to high partial waves $l$, but the calculations
slow down considerably for heavy atoms due to two reasons. First,
the $1s$ core orbitals become smaller and smaller going deeper in
the periodic table, thus requiring smaller length scales to be represented
in the partial wave expansion. Second, heavier atoms are also bigger,
which makes the bond lengths larger, further decreasing the angular
size of the core orbitals. Going to heavier atoms will also require
the treatment of relativistic effects, which have not been discussed
in the present work.

While the Coulomb matrix can be formed in little time even in large
calculations, the exchange contributions in both HF and DFT scale
as the cube of the number of basis functions, $N^{3}$. However, as
the program is still quite new, it is quite likely that these problems
can be circumvented via optimizations to the code and/or further parallellization.

\subsection{Post-HF approaches\label{subsec:Post-HF-approaches}}

Although fully numerical calculations on atoms and diatomic molecules
afford arbitrary precision for atoms and diatomic molecules within
Hartree--Fock (HF) or density functional theory (DFT), the accuracy
of these methods themselves is insufficient for many applications.
More accurate estimates can be obtained by switching to a post-HF
level of theory, such as multiconfigurational theory,\citep{Hinze1967,Hinze1973}
density matrix renormalization group theory,\citep{White1992} or
coupled-cluster theory.\citep{Cizek1966} However, going to the post-HF
level causes further complications.

As is commonly known from Gaussian-basis calculations, the basis set
convergence of the correlation energy is very slow.\citep{Wilson1996,Wilson1996b,Helgaker1997a,Wilson1997a,Halkier1998,Moncrieff1996}
In fact, the problem appears already for two-electron systems,\citep{Schwartz1962}
\emph{e.g.} in configuration interaction calculations on the $1s^{2}$
state of He as discovered in the late 1970s in seminal work by Silverstone
and coworkers.\citep{Silverstone1978,Carroll1979} Because of this,
even atomic fully numerical calculations have to rely on extrapolation
approaches.\citep{Flores2004,Flores2006,Flores2008} Although extrapolation
techniques are commonly used even for SCF energies in the context
of LCAO calculations, as the SCF energy affords exponential convergence
such approaches are usually not necessary except for highly accurate
benchmark calculations. Regardless of the used basis set, the key
to obtaining accurate post-HF energetics, however, is the extrapolation
to the CBS limit, which has been studied extensively in the literature
using a wealth of extrapolation formulas.\citep{Nyden1981,Hill1985,Feller1992,Feller1993,Peterson1994,Martin1996,Wilson1997a,Halkier1998,Halkier1999,Balabanov2003,Bytautas2004,Jensen2005a,Dixon2005,Peterson2005,Schwenke2005,Karton2006,Karton2006a,Bakowies2007,Lee2007,Varandas2007,Hill2009,Feller2011,Peterson2012,Feller2013,Okoshi2015,Mezei2015,Boschen2017}

Moreover, although it is extremely simple to obtain large sets of
virtual orbitals for both atoms and diatomic molecules,\citep{Lehtola2019a,Lehtola2019b}
it is unlikely that they can be fully used in correlation treatments.
Even calculating and storing the two-electron integrals, which is
a rank-4 tensor, may rapidly become challenging in large numerical
basis sets. For instance, the number of operations in configuration
interaction (CI) and coupled-cluster theory scales as $O(o^{n}v^{n+2})$,
where $o$ and $v$ are the the number of occupied and virtual orbitals,
respectively, where $n$ is the maximum level of substitutions included
in the calculation. Taking the $n=2$ case for doubles, the $v^{4}$
scaling quickly makes calculations infeasible, meaning that only a
small subset of virtual obitals can be included in the treatment.
If all excitations are included, as in the full CI approach, the scaling
becomes even more prohibitive: per common knowledge, one is limited
to studying problem sizes of at most 20 electrons in 20 orbitals due
to the exponential scaling algorithm. While the density matrix renormalization
group\citep{White1992} approach is a more efficient parametrization
of the correlation problem, it too is exponentially scaling in general,
and is limited in such cases to problem sizes of about 50 electrons
in 50 orbitals.\citep{Olivares-Amaya2015}

Although post-HF calculations on atoms and diatomic molecules have
been reported by several authors in the literature, as discussed above,
specialized algorithms for atoms and diatomic molecules have fewer
advantages over general three-dimensional programs at post-HF levels
of theory, since a large set of virtual orbitals becomes necessary.
It may then be less important to be able to treat the mean-field problem
to arbitrary precision, when significant errors may still be made
in the correlation treatment due the intrinsic limitations therein
caused by the computational scaling.

Despite this slightly disconcerting analysis, there is still room
for specialized calculations on atoms and diatomic molecules. In cases
where HF or DFT is sufficient, a specialized numerical approach is
more cost-effective than a general one. Furthermore, the development
and verification of novel, general three-dimensional programs benefits
from extremely accurate benchmark data which is achievable through
the specialized atomic and diatomic approaches. The specialization
of post-HF to the case of atoms or diatomic molecules may allow for
a more efficient use of symmetries as well, enabling one to perform
slightly larger calculations than within a completely general approach.

\section*{Funding information}

This work has been supported by the Academy of Finland (Suomen Akatemia,
Luonnontieteiden ja Tekniikan Tutkimuksen Toimikunta) through project
number 311149.

\section*{Acknowledgments}

I thank Dage Sundholm and Jacek Kobus for discussions, and Dage Sundholm,
Lukas Wirz, Jacek Kobus, and Pekka Pyykkö for comments on the manuscript.

\section*{Appendix A: orbitals in linear symmetry}

To ensure general readability of the manuscript, I remind here that
orbitals in linear molecules with $m=0$, $m=\pm1$, $m=\pm2$, and
$m=\pm3$ are commonly known as $\sigma$, $\pi$, $\delta$, and
$\varphi$ orbitals, respectively. Orbitals beyond $\varphi$ are
not necessary for SCF calculations in the known periodic table, although
they may be necessary for post-HF approaches. A fully filled $\sigma$
orbital can fit two electrons, whereas fully filled $\pi$, $\delta$,
and $\varphi$ orbitals can fit four electrons, as both the $m=|m|$
and $m=-|m|$ subshells fit two electrons.

It is worthwhile to point out here that a diatomic $\sigma$ orbital
includes the $m=0$ components from all atomic orbitals: in addition
to the $ns$ orbitals, contributions may also arise from atomic $np_{z}$,
$nd_{z^{2}}$, $nf_{z^{3}}$, \emph{etc.} orbitals which describe
polarization effects in LCAO calculations; similar remarks can be
also made about the $\pi$, $\delta$, and $\varphi$ orbitals.

The other way around, an atomic $ns$ orbital yields a $\sigma$ orbital,
whereas an atomic $np$ orbital yields one $\sigma$ and two $\pi$
orbitals, corresponding to the $m=0$ and $m=\pm1$ components of
the function. Similarly, atomic $nd$ orbitals yield one $\sigma$,
two $\pi$, and two $\delta$ functions, whereas atomic $nf$ orbitals
yield one $\sigma$, two $\pi$, two $\delta$, and two $\varphi$
orbitals.

\section*{Author Biography}

\includegraphics[height=3cm]{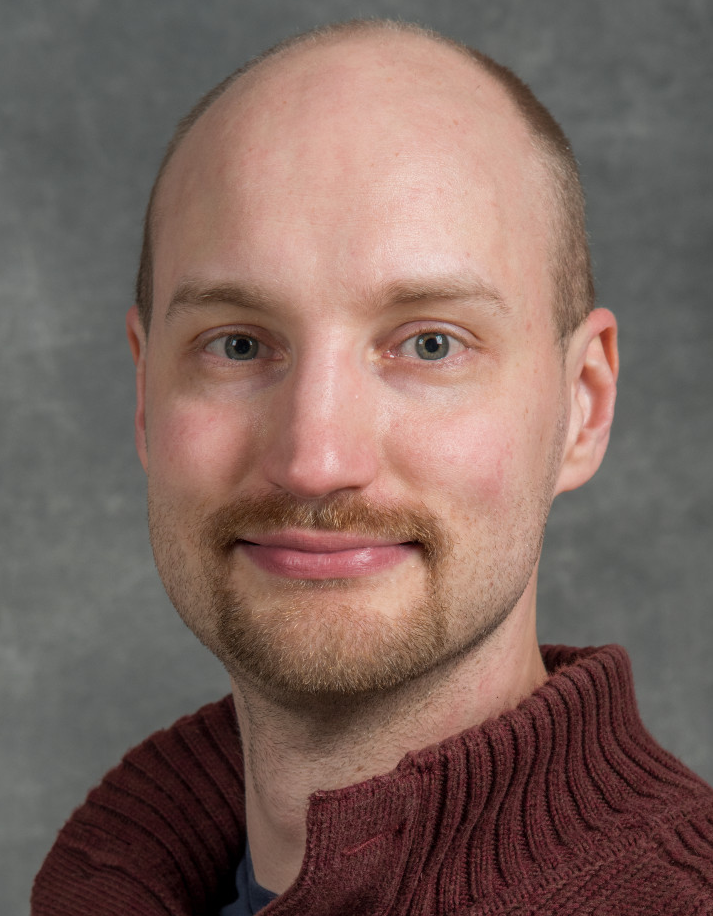}

Susi Lehtola completed his PhD in theoretical physics in 2013 at the
University of Helsinki. After a one-year postdoc with Hannes Jónsson
at Aalto University and a three-year postdoc with Martin Head-Gordon
at University of California, Berkeley, he was awarded the Löwdin postdoctoral
award at the Sanibel meeting 2017, as well as a three-year Academy
of Finland Postdoctoral Researcher scholarship in 2017 for methods
development at the University of Helsinki.

\bibliographystyle{achemso}
\bibliography{citations}

\end{document}